\documentclass{article}
\usepackage{arxiv}

\usepackage[utf8]{inputenc} % allow utf-8 input
\usepackage[T1]{fontenc}    % use 8-bit T1 fonts
\usepackage[hidelinks]{hyperref}     % hyperlinks
\usepackage{url}            % simple URL typesetting
\usepackage{booktabs}       % professional-quality tables
\usepackage{amsfonts}       % blackboard math symbols
\usepackage{nicefrac}       % compact symbols for 1/2, etc.
\usepackage{microtype}      % microtypography
\usepackage{lipsum}
\usepackage{bm}
\usepackage{graphicx}
\usepackage{subcaption}
\usepackage{amsmath}
\usepackage{placeins}
\graphicspath{ {./images/} }

\title{Generalized Eshelby's inclusion and inhomogeneity problems for transient heat transfer}

\author{
 Chunlin Wu \\
  Shanghai Institute of Applied Mathematics and Mechanics\\
  Shanghai University\\
  Shanghai, 200044 \\
  \texttt{chunlinwu@shu.edu.cn} \\
   \And
 Zhenhua Wei \\
 Department of Ocean Science and Engineering \\
 Southern University of Science and Technology \\
 Shenzhen, 518055 \\
  \texttt{weizh@sustech.edu.cn} \\
  \And
 Huiming Yin \\
  Department of Civil Engineering and Engineering Mechanics\\
  Columbia University\\
  New York, NY, 10027 \\
  \texttt{yin@civil.columbia.edu} \\
}

\begin{document}
\maketitle
\begin{abstract}
Eshelby's inclusion problems have been generalized to arbitrary shape of polygonal, polyhedral, and ellipsoidal inclusions embedded in an infinite isotropic domain under transient heat transfer, and Eshelby's tensors have been analytically derived to evaluate disturbed thermal fields caused by inclusions with a polynomial-form eigen-field. Transformed coordinates are applied to arbitrarily shaped inclusions for domain integrals of transient fundamental solutions. This formulation is for general transient heat transfer, and it can recover classic Eshelby's tensor for the ellipsoidal subdomain with explicit expression for the spherical domain in the steady state, Michelitsch's solution in the harmonic state, and recent solution in the transient state. The formulae for a polyhedral inclusion is verified by comparison to closed-form solutions of a spherical inclusion when the sphere is divided into many polyhedrons. The discontinuity of domain integrals for Eshelby's tensor are investigated the temporal effects are elaborated. The generalized formulation for Eshelby's problems enables the simulation and modeling of particulate composites containing inhomogeneities of various shapes for steady-state, harmonic and transient heat transfer in both two- and three-dimensional space through the equivalent inclusion method.
\end{abstract}

\keywords{Transient heat transfer \and Polynomial-form eigen-fields \and Inclusion and inhomogeneity \and Equivalent inclusion method \and Eshelby's tensors }

\section{Introduction}
Heat transfer processes have been widely used in different engineering applications, such as building envelopes in civil engineering \cite{Hailu2021}, physical and thermal barriers in aerospace engineering \cite{Kumar2013}, and combustion engines in automobile engineering \cite{Ram2017}. The design and applications of these engineering materials and systems can usually be reduced to problems in determining thermal fields, such as temperature and heat flux. Despite the significance of materials design for thermal engineering, there is a lack of research attempts to develop analytical methods to elucidate the roles of inclusions/inhomogeneities in affecting heat transfer in the literature, especially in the case of transient heat conduction in solids.

The investigation of disturbances caused by inclusion/inhomogeneity can be traced back to the 1950s. Dealing with an ellipsoidal inhomogeneity is embedded in an infinite isotropic medium, Eshelby \cite{Eshelby_1957} proposed the celebrated equivalent inclusion method (EIM) which replaces the inhomogeneity by the matrix and a continuously distributed eigenstrain. Thanks to Eshelby's significant contribution on inhomogeneity disturbances, the EIM has been extended in several multi-physical problems, including magneto-elasticity\cite {yin2006magneto}, piezoelectricity \cite{Pan_2004, Zeng2003} and thermal analysis \cite{Hatta1986,Wu2022_JMPS}. The intrinsic feature of the EIM is to determine an accurate or at least approximated eigen-field that satisfies a group of linear equations, namely equivalent conditions. Solving equivalent conditions generally requires domain integrals of fundamental solutions, namely Eshelby's tensor; therefore, extensive efforts have been devoted to exploring Eshelby's tensor. 

% Eshelby's tensors on static problems:
The original elastic Eshelby's tensors of ellipsoidal inclusions are composed of the harmonic and biharmonic potentials, which were first derived by Dyson \cite{Dyson1891}. Subsequently, Jin et al. derived a complete set of analytical Eshelby's tensors for ellipsoidal inclusions on elastostatics \cite{Jin2016}. As most works focused on ellipsoidal/elliptical inclusion/inhomogeneities, the lack of Eshelby's tensors for arbitrarily-shaped limited EIM's applications. Hence, efforts have been shifted to investigate disturbance caused by arbitrarily shaped inclusions \cite{Nozaki1997,Ru1999}. Using Waldvogel's work on Newtonian potential \cite{Waldvogel1979}, Rodin \cite{Rodin1996} first derived closed-form Eshelby's tensor for polygonal and polyhedral inclusions with a uniform source field using transformed coordinates and area/surface integral can be further reduced to contour and surface integral through the divergence theorem. Rodin's work \cite{Rodin1996} paved the way for subsequent works by Gao's group on polyhedrons \cite{Gao2012} and polygons \cite{Liu2013}, which involved Mindlin's strain gradient theory in considering eigenstrain variations for size effects. To shorten lengthy expressions and avoid trivial coordinate transformation, Trotta's group \cite{Trotta2017, Trotta2018} proposed Eshelby's tensor on polygonal and polyhedral inclusions, which were directly obtained from vertices. 

Note that the above papers considered a constant eigenstrain for inclusion problems only. Eshelby's tensors cannot reflect the nonuniform distribution of eigen-fields frequently used in inhomogeneity problems, especially for angular particles or many particles with pair interactions and boundary effects. Recently, Yin's group \cite{Wu2021_jam_polygonal, Wu2021_polyhedral} proposed a closed-form Eshelby's tensor for a polygonal and polyhedral inclusion with polynomial-form eigenstrain expanded at its center. Their results revealed that the higher-order terms could significantly improve the accuracy of EIM, which was also confirmed in a subsequent extension to a two-dimensional bimaterial thermoelastic problem \cite{Wu2023_ijss}. Moreover, Wu and Yin \cite{Wu2024_JMPS} have rigorously proved the singular eigenstrain around triangular inhomogeneities even subjected to uniform far-field loads. However, only a few studies have considered time-dependent loading conditions. 

To investigate dynamic behaviors caused by inclusions, Fu and Mura \cite{Fu1982} derived a series-form domain integral over an ellipsoidal domain, which expanded at the distance of source and field points, respectively. Subsequently, the same authors \cite{Fu1983} employed the series-form domain integrals to solve inhomogeneity problems considering polynomial-form distributed eigenstrain and eigenbodyforce. Mikata and Nemat-Nasser \cite{Mikata1990} focused on specific engineering applications of spherical inclusions and derived closed-form domain integrals. Michelitsch et al. \cite{Michelitsch2003} utilized the convolution property of the Fourier space to obtain a general integral scheme for domain integrals, in which the domain integrals for ellipsoids are transformed into a two-dimensional (2D) surface integral for general ellipsoids and spheroids with exterior field points, a one-dimensional (1D) integral for spheroids with interior field points, and the closed-form domain integrals for spheres \cite{Mikata1990}. Michelitsch \cite{Michelitsch2003_shell} also simplified the 2D surface integrals as a one-dimensional integral using series expansion with the frequency parameter $\beta$. Subsequently, the authors extended their works on Helmholtz's potential to retarded potential of time-domain \cite{Wang2005, Michelitsch2005} on the arbitrary source and following Eshelby's problems, respectively. In 2018, Delfani and Sajedipour \cite{Delfani2018} combined the strain gradient theory in elastodynamics on spherical inclusion with an integral-form solution. Our recent work investigated the harmonic heat transfer problem for an ellipsoidal inhomogeneity, which derives closed-form domain integral of harmonic Green's function, Eshelby's tensor, and then applied it to the equivalent inclusion method for harmonic heat transfer \cite{Wu2024-prsa}. The above studies paved the way for the present investigation of transient heat conduction. 

% Scope of the paper
This paper extends the recent work in harmonic heat transfer \cite{Wu2024-prsa} to transient heat transfer analysis, and generalizes the concept of Eshelby's tensors for domain integrals of both both eigen-heat-source and eigen-temperature gradient in the polynomial form. Different shapes of inclusions, including polygonal, polyhedral, and ellipsoidal inclusions are considered, so that the corresponding inhomogeneity problems can be solved with Eshelby's EIM. Note that Eshelby's tensors depend on Green's functions and geometry of inclusions. It is the first time in the literature to show the Eshelby's tensors in some loading scenarios and inclusion shapes through this generalized framework, which can be reduced into some special cases, such as elastostatic and steady-state heat conduction problems. 

In the following, Section 2 reviews the heat equation with a fundamental solution for transient heat transfer. Section 3 derives the domain integrals of fundamental solutions over arbitrarily-shaped polygonal and polyhedral inclusions. Section 4 derives the domain integrals for ellipsoidal inclusions through Fourier analysis and provides the closed-form solution for spherical inclusion. Section 5 verifies the polyhedral domain integrals by comparing closed-form domain integrals over a spherical domain, and the domain integrals by the Fourier transform are verified through the comparison to cuboid domain integrals. And the equivalent inclusion method is conducted over the spheroidal inhomogeneity, which was verified by the finite element method. Finally, some conclusive remarks are presented in Section 6. 

\section{Formulation}

\subsection{Governing equation and fundamental solution}
The heat equation governs heat transfer in a homogeneous medium. Given a heat source at source point $\textbf{x}'$ at time $t'$, the heat equation is written as follows:

% heat equation
\begin{equation}
    C_p \frac{\partial u(\textbf{x}, t)}{\partial t} = K \nabla^2 u(\textbf{x}, t) + Q (\textbf{x}', t') \quad \text{or} \quad \frac{\partial u(\textbf{x}, t)}{\partial t} = \alpha \nabla^2 u(\textbf{x}, t) + \frac{Q(\textbf{x}', \textbf{t}')}{C_p}
\label{eq:govern_eqn}
\end{equation}
where $\textbf{x}$ and $\textbf{x}'$ represent field and source points, respectively; $\alpha = \frac{K}{C_p}$ refers to the thermal diffusivity; $u$ is the temperature field; $t$ and $t'$ stand for current time and time when the source is applied, respectively; and $Q$ is the distributed heat source. For a partial differential equation with infinite boundary condition (far-field zero temperature), the temperature at any field point $\textbf{x}$ can be expressed in terms of the fundamental solution and applied heat source, 

% field point temperature
\begin{equation}
    u(\textbf{x}, t) = \frac{1}{C_p} \int_{t_i}^{t_f} \int_{\Omega} G(\textbf{x}, \textbf{x}', t, t') Q(\textbf{x}', t') d\textbf{x}' d t'
    \label{eq:temp}
\end{equation}
where $t_i$ and $t_f$ describe the starting and ending time of a heat source ($t_i<t_f\leq t$), $G(\textbf{x}, \textbf{x}', t, t')$ is the fundamental solution to the heat equation. The fundamental solutions or Green's functions for 2D and 3D problems are slightly different as follows \cite{Crank1947}, 
% fundamental solution
\begin{equation}
    G(\textbf{x}, \textbf{x}', t, t') = [4 \pi \alpha (t - t')]^{\frac{-n}{2}} \exp \left[ \frac{-|\textbf{x} - \textbf{x}'|^2}{4 \alpha (t - t')} \right] H(t - t')
    \label{eq:fund_time}
\end{equation}
where $n = 2, 3$ is for 2D and 3D problems, respectively; $H(t - t')$ is the Heaviside function. Note that Eq. (\ref{eq:fund_time}) relates heat source and temperature for transient heat transfer and depends on $\textbf{x}-\textbf{x}'$ and $t-t'$ only, although four variables are listed as independent for the ease of notation in integrals subsequently. 

When the heat source follows a harmonic function with an excitation $\omega$, as $u(\textbf{x}, t) = \overline{u}(\textbf{x}) \exp[-i \omega t]$, the heat equation can be changed to a modified Helmholtz's equation \cite{Wu2024-prsa}. The temporal and spatial variables can be separated, and the fundamental solution and domain integrals can be conducted in the spatial domain, which is elaborated in \ref{appendix:harmonic}.

\subsection{Disturbance caused by eigen-fields of an inclusion}
The following subsection considers disturbance caused by continuously distributed source fields within one inclusion. Following our recent work in harmonic heat transfer \cite{Wu2024-prsa}, two eigen-fields, eigen-heat-source (EHS, $Q^*$) and eigen-temperature-gradient (ETG, $u_i^*$), are employed to simulate material mismatch between the matrix and inhomogeneities on specific heat and thermal conductivity, respectively. The original heat equation in Eq. (\ref{eq:govern_eqn}) is rewritten as follows:

\begin{equation}
    C_p u_{,t}(\textbf{x}, t) = K [u_{,ii}(\textbf{x} , t) - u_{i,i}^{*}(\textbf{x} , t)]  +Q(\textbf{x} , t) - Q^{*}(\textbf{x} , t) 
\label{eq:govern_time2}
\end{equation}
where eigen-fields are continuous over the inclusion. Therefore, they can be written in terms of polynomials by the Taylor's expansion \cite{Mura1987} as follows: 
% polynomial-form eigen-fields
\begin{equation}
\begin{aligned}
    & Q^* (\textbf{x}, t) = Q^{0*}(t) + Q^{1*}_{p}(t) (x_p - x_p^{C}) + Q^{2*}_{pq}(t)  (x_p - x_p^{C}) (x_q - x_q^{C}),  \quad \textbf{x} \in \Omega \\
    & u_i^{*}(\textbf{x}, t) = u_{i}^{0*}(t) +  u^{1*}_{ip}(t) (x_p - x_p^{C}) + u^{2*}_{ipq}(t)  (x_p - x_p^{C}) (x_q - x_q^{C}), \quad \textbf{x} \in \Omega
\end{aligned}
    \label{eq:poly_eigen}
\end{equation}
where eigen-fields only exist within the inclusion $\Omega$; $\textbf{x}^{C}$ is the center of $\Omega$; superscripts $0*, 1*$ and $2*$ stand for the uniform, linear and quadratic polynomial terms of the eigen-fields. Mathematically, the polynomials can be obtained by the Taylor expansion of the eigen-fields referred to the particle center with their derivatives matching with the parameters at each order of $0*, 1*$ and $2*$. And this paper considers up to quadratic order \cite{Mura1987}. Treating $-K u_{i,i}^*(\textbf{x}, t) - Q^*(\textbf{x}, t)$ as an artificial heat source, the disturbance by eigen-fields can be evaluated through domain integrals of fundamental solutions over $\Omega$ in the following methods: 

For the effect of EHS $ Q^* (\textbf{x}, t)$, the Green's function is directly used in the integral as:

\begin{small}
\begin{equation}
\begin{split}
    u(\textbf{x}, t) & = -\frac{1}{C_p} \int_{t_i}^{t_f} \thinspace \Big\{ Q^{0*}(t') \int_{\Omega} G(\textbf{x}, \textbf{x}', t, t') \thinspace d \textbf{x}' + Q^{1*}_{p}(t') \int_{\Omega} G(\textbf{x}, \textbf{x}', t, t') (x'_p - x_p^{C}) \thinspace d \textbf{x}' \\
    & \thinspace \thinspace + Q^{2*}_{pq}(t') \int_{\Omega} G(\textbf{x}, \textbf{x}', t, t') (x'_p - x_p^{C}) (x'_q - x_q^{C}) \thinspace d\textbf{x}' \Big\} dt'
\end{split}
    \label{eq:disturbance_heatsource}
\end{equation}
\end{small}
For the effect of ETG $u_{i}^*(\textbf{x}, t)$, Gauss' theorem can be applied to yield the temperature variation as:
\begin{equation}
\begin{split}
    & u(\textbf{x}, t) = \alpha \int_{t_i}^{t_f} \Big\{ \int_{\Omega} G_{,i'}(\textbf{x}, \textbf{x}', t, t') {u}_i^{*} (\textbf{x}', t')\thinspace d\textbf{x}' -\int_{\partial\Omega} G(\textbf{x}, \textbf{x}', t, t') u_i^{*}(\textbf{x}', t') n_i \thinspace d\textbf{x}' \Big\} dt' \\
    %
    % & = \alpha \int_{t_i}^{t_f} \Big\{ \int_{\Omega} G_{,i'}(\textbf{x}, \textbf{x}', t, t') \left[ {u}_i^{0*}(t') + {u}_{ip}^{1*}(t') (x'_p - x_p^{C}) + {u}_{ipq}^{2*}(t') (x'_p - x_p^{C}) (x'_q - x_q^{C}) \right] \thinspace d\textbf{x}' \Big\} dt' \\ 
    % %
    & = \alpha \int_{t_i}^{t_f} \Big\{ \int_{\Omega} \left[ {u}_i^{0*}(t') \int_{\Omega} G_{,i'}(\textbf{x}, \textbf{x}', t, t') \thinspace d \textbf{x}' + {u}_{ip}^{1*}(t') \int_{\Omega} G_{,i'}(\textbf{x}, \textbf{x}', t, t') (x'_p - x_p^{C}) \thinspace d \textbf{x}' \right. \\ & \left. + {u}_{ipq}^{2*}(t') \int_{\Omega} G_{,i'}(\textbf{x}, \textbf{x}', t, t')  (x'_p - x_p^{C}) (x'_q - x_q^{C}) \thinspace d\textbf{x}' \right]  \Big\} dt'
\end{split}
    \label{eq:disturbance_ETG}
\end{equation}
where $\partial \Omega$ represents the interface between the inclusion and the matrix; the surface integral is zero, as piece-wisely continuous eigen-fields vanish on the outer boundary surface. Note that Eshelby's tensor was originally defined to predict the disturbed strain field caused by a uniform eigenstrain on an ellipsoidal inclusion \cite{Eshelby_1957,Eshelby_1959,Mura1987} , and the eigenstrain can be considered as a counterpart of ETG in thermal analysis. For the steady-state, Eshelby's tensor is constant in the inclusion \cite{Yin_iBEM}. For transient heat transfer problems, the concept of Eshelby's tensor is generalized to the domain integrals of both ETG and EHS in the polynomial form.

Superposing disturbance caused by EHS and ETG, the temperature at any field point $\textbf{x}$ and time $t$ can be acquired as the spatio-temporal domain integrals of fundamental solutions multiplied by polynomial-form source fields, 

% temperature - superposition
\begin{small}
\begin{equation}
    u(\textbf{x}, t) = \underbrace{ \left( Q^{0*} \int_{t_i}^{t_f} L d t'  + Q_p^{1*} \int_{t_i}^{t_f} L_{p} d t' + Q_{pq}^{2*} \int_{t_i}^{t_f} L_{pq} dt' \right)}_{\text{by eigen-heat-source}} + \underbrace{ \left( u_{i}^{0*} \int_{t_i}^{t_f} D_{i} dt'   + u_{ip}^{1*} \int_{t_i}^{t_f} D_{ip} dt'  + u_{ipq}^{2*} \int_{t_i}^{t_f} D_{ipq} dt'  \right)}_{\text{by eigen-temperature-gradient}}
    \label{eq:super_disturb}
\end{equation}
\end{small}
where $L, L_p, L_{pq}$ and $D_i, D_{ip}, D_{ipq}$ are spatial Eshelby's tensors for polynomial-form of EHS and ETG, respectively, as follows: 
% definition of Eshelby's tensors
\begin{small}
\begin{equation}
    \begin{aligned}
    & L_{pq...}(\textbf{x}, \textbf{x}^C, t, t') =- \frac{1}{C_p} \int_{\Omega} (x_p' - x_p^C) (x_q' - x_q^C)... G(\textbf{x}, \textbf{x}', t, t')\thinspace d\textbf{x}' \\
    & D_{ipq...}(\textbf{x}, \textbf{x}^C, t, t') = \alpha \int_{\Omega} (x_p' - x_p^C) (x_q' - x_q^C)... G_{,i'}(\textbf{x}, \textbf{x}', t, t')\thinspace d\textbf{x}' = -\alpha \int_{\Omega} (x_p' - x_p^C) (x_q' - x_q^C)... G_{,i}(\textbf{x}, \textbf{x}', t, t')\thinspace d\textbf{x}'  \\
    \end{aligned}
    \label{eq:Eshelby_tensor_spatial}
\end{equation}
\end{small}
in which $\textbf{x}^C$ is the center of the inclusion domain; the subscripts $p, q, \cdots$ mean the order of polynomial, i.e., null for uniform, $p$ for linear, and $pq$ for quadratic; $G_{,i'}(\textbf{x}, \textbf{x}', t, t') = -G_{,i}(\textbf{x}, \textbf{x}', t, t')$ is valid for full-space fundamental solutions. Without the loss of any generality, this paper sets the origin at $\textbf{x}^C = \textbf{0}$. Therefore, spatial Eshelby's tensors can be further written as $L_{pq...}(\textbf{x}, \textbf{x}^C, t, t') = L_{pq...}(\textbf{x}, t, t')$ and $D_{ipq...}(\textbf{x}, \textbf{x}^C, t, t') = D_{ipq...}(\textbf{x}, t, t')$. 

Eq. (\ref{eq:super_disturb}) requires time integral to derive the time Eshelby's tensor. The spatial variation of eigen-fields is regarded as a polynomial, and we assume that between two adjacent time stations, i.e., $t_f, t_{f-1}$, the time-dependent quantities are constant. For instance, polynomial-form eigen-fields in Eq. (\ref{eq:poly_eigen}) can be written as \cite{Wu_IJSS_2025}, 

\begin{equation}
\begin{aligned}
    & Q^{*}(\textbf{x}, t) = \sum_{f = 1}^{N_f} H\left[ (t - t_{f-1} ) (t_f - t)\right]  \Big( Q^{f0*} +  Q_{p}^{f1*} x_p + Q_{pq}^{f2*} x_p x_q \Big) \\
    & u^{*}(\textbf{x}, t) = \sum_{f = 1}^{N_f} H\left[ (t - t_{f-1} ) (t_f - t)\right]  \Big( u_i^{f0*} +  u_{ip}^{f1*} x_p + u_{ipq}^{f2*} x_p x_q \Big)
\end{aligned}
    \label{eq:poly_time_eigen}
\end{equation}
where the Heaviside function $H\left[ (t - t_{f-1} ) (t_f - t)\right]$ indicates step-wise time-dependent eigen-fields with nonzero between $(t_{f-1},t_f)$; $N_f$ is the number of time steps; superscripts $f0*, f1*$ and $f2*$ represent uniform, linear, and quadratic coefficients within the time window $[t_{f-1}, t_f]$. The assumption on constant quantities over a step can be extended to more complicated shape functions, such as linear, quadratic polynomials. For instance, although Eq. (\ref{eq:3D_time}), Eq. (\ref{eq:stokes_time}), and Eqs. (\ref{eq:sphere_time_uni}-\ref{eq:sphere_time_qua}) does not consider time varying quantities, the time integrals (multiplying by linear and quadratic time variables) can be conducted similarly. Subsequently, the disturbed temperature and heat flux can be written as,

\begin{equation}
\begin{split}
    u(\textbf{x}, t) = & \sum_{f=1}^{N_f} \left\{ Q^{f0*} \left( \overline{L}^f(\textbf{x}, t)  + Q_p^{f1*} \overline{L}^f_p(\textbf{x}, t) + Q_{pq}^{f2*} \overline{L}^f_{pq}(\textbf{x}, t) \right) \right. \\ & 
    \left.
    + \left( u_i^{f0*} \overline{D}_i^f(\textbf{x}, t)  + u_{ip}^{f1*} \overline{D}^f_{ip}(\textbf{x}, t) + u_{ipq}^{f2*} \overline{D}^f_{ipq}(\textbf{x}, t) \right) \right\}
\end{split}
    \label{eq:super_disturb_time}
\end{equation}

\begin{equation}
\begin{split}
    q_m(\textbf{x}, t)  = & -K \sum_{f=1}^{N_f} \left\{ Q^{f0*} \left( \overline{L}^f_{,m}(\textbf{x}, t)  + Q_p^{f1*} \overline{L}^f_{p,m}(\textbf{x}, t) + Q_{pq}^{f2*} \overline{L}^f_{pq,m}(\textbf{x}, t) \right) \right. \\ & \left. + \left( u_i^{f0*} \overline{D}_{i,m}^f(\textbf{x}, t)  + u_{ip,m}^{f1*} \overline{D}^f_{ip}(\textbf{x}, t) + u_{ipq}^{f2*} \overline{D}^f_{ipq,m}(\textbf{x}, t) \right) \right\}
\end{split}
    \label{eq:flux}
\end{equation}
where the disturbed temperature is a discretized convolution of the fundamental solution over the inclusion and time domain. $\overline{L}^f, \overline{L}^f_{p}, \overline{L}^f_{pq}$ and $\overline{D}^f_{i}, \overline{D}^{f}_{ip}, \overline{D}^f_{ipq}$ are time Eshelby's tensors for polynomial-form of EHS and ETG within the time station $[t_{f-1}, t_f]$, respectively. Similarly to the spatial Eshelby's tensors in Eq. (\ref{eq:Eshelby_tensor_spatial}), they are written as: 

% definition of Eshelby's tensors
\begin{equation}
    \overline{L}^f_{pq...}(\textbf{x}, t) = \int_{t_{f-1}}^{t_f} L_{pq...}(\textbf{x}, t, t') \thinspace d t' \quad \text{and} \quad \overline{D}^f_{ipq...}(\textbf{x}, t) = \int_{t_{f-1}}^{t_f} D_{ipq...}(\textbf{x}, t, t') \thinspace d t'
    \label{eq:Eshelby_tensor_time}
\end{equation}
Note that although time steps are not required to be equal, using the same time step can significantly accelerate the time convolution process, which have been explained in \cite{Gupta1995, Brebbia1984}. For instance, if one is seeking disturbed temperature at the time $t_1, t_2$ with unequal time steps, (i) time Eshelby's tensors should be calculated between time $[t_0, t_1]$ for disturbed temperatures at $t_1$, and; (ii) time Eshelby's tensors should be calculated between time $[t_1, t_2]$ for disturbed temperatures at $t_2$, which consider influences from source fields existing within two time intervals, respectively. When $t_1 - t_0$ is not equivalent to $t_2 - t_1$, the evaluation of disturbed temperature at $t_2$ requires two evaluations of coefficients. However, this issue can be handled by using equal time steps, so that only one evaluation is required to calculate $\overline{L}^{1}(\textbf{x}, t)$ and $ \overline{D}^{1}(\textbf{x}, t)$. Therefore, this paper employs equal time steps. 

\subsection{Pre-processing of the polynomial-form directional source terms}
Since eigen-fields are expressed through the Taylor series expansion at the center (origin), they generally contain directional source terms, such as $x_p'$. The existence of such directional source terms makes it difficult to evaluate domain integrals. Instead of direct integration, our recent work \cite{Wu2024-prsa} proposed to rewrite directional source terms as partial derivatives of a higher-order potential function. It is essential to rewrite the integral variable $\textbf{x}'$ referred to the field point $\textbf{x}$ as follows

\begin{equation}
    x_p' = (x_p' - x_p) + x_p \quad \text{and} \quad
    x_p' x_q' = (x_p' - x_p) (x_q' - x_q) + x_p (x_q'-x_q) + (x_p' - x_p) x_q + x_p x_q 
\label{eq:treat_poly}
\end{equation}

Therefore, Eq. (\ref{eq:Eshelby_tensor_spatial}) can be rewritten in terms of the partial derivatives of the fundamental solution \cite{Wu2024-prsa}. For example, the linear and quadratic source terms multiplied fundamental solutions can be rewritten as follows:

\begin{equation}
\begin{aligned}
     G_p(\textbf{x}, \textbf{x}', t, t') &= x_p' G(\textbf{x}, \textbf{x}', t, t') = 2 \alpha (t - t') G_{,p}(\textbf{x}, \textbf{x}', t, t') + x_p G(\textbf{x}, \textbf{x}', t, t') \\
     G_{pq}(\textbf{x}, \textbf{x}', t, t') &= x_p' x_q' G(\textbf{x}, \textbf{x}', t, t') = (2 \alpha (t - t') )^2 G_{,pq}(\textbf{x}, \textbf{x}', t, t') \\ & + 2 \alpha (t - t') \Big( \delta_{pq} G(\textbf{x}, \textbf{x}', t, t') + x_p G_{,q}(\textbf{x}, \textbf{x}', t, t') + x_q G_{,p}(\textbf{x}, \textbf{x}', t, t') \Big) + x_p x_q G(\textbf{x}, \textbf{x}', t, t')
\end{aligned}
\label{eq:poly_before_int}
\end{equation}
which can significantly simplify the derivation of the generalized Eshelby's tensor. In such a case, domain integrals of linear and quadratic source terms can be determined by modifying the uniform spatial Eshelby's tensor $L(\textbf{x}, t, t')$. Specifically, the linear and quadratic domain integrals can be expressed as,

\begin{equation}
    \begin{aligned}
    & L_p(\textbf{x}, t, t') = 2 \alpha (t - t') L_{,p}(\textbf{x}, t, t') + x_p L(\textbf{x}, t, t') \\
    & L_{pq}(\textbf{x}, t, t') = (2 \alpha (t - t'))^2 L_{,pq}(\textbf{x}, t, t') + 2 \alpha (t - t') \Big( \delta_{pq} L(\textbf{x}, t, t') + x_p L_q(\textbf{x}, t, t') + x_q L_p(\textbf{x}, t, t')\Big) + x_p x_q L(\textbf{x}, t, t')
    \end{aligned}
     \label{eq:L_spatial}
\end{equation}

Based on Eq. (\ref{eq:Eshelby_tensor_spatial}), spatial Eshelby's tensors $L_{pq...}$ and $D_{ipq...}$ are related through a further partial differentiation and a multiplier $K$, therefore, spatial tensors $D_{ip}, D_{ipq}$ for ETG can be written as: 

\begin{equation}
    \begin{aligned}
    & D_{ip}(\textbf{x}, t, t') = -K \Big\{ 2 \alpha (t - t') L_{,ip}(\textbf{x}, t, t') + \delta_{ip} L(\textbf{x}, t, t') + x_p L_{,i}(\textbf{x}, t, t') \Big\} \\
    & D_{ipq}(\textbf{x}, t, t') = -K \Big\{ (2 \alpha (t - t'))^2 L_{,ipq}(\textbf{x}, t, t') + 2 \alpha (t - t') \Big( \delta_{pq} L_{,i}(\textbf{x}, t, t') + \delta_{ip} L_q(\textbf{x}, t, t') + \delta_{iq} L_p(\textbf{x}, t, t') \\ & \qquad \qquad \qquad + x_p L_{q,i}(\textbf{x}, t, t') + x_q L_{p,i}(\textbf{x}, t, t') \Big) + (\delta_{ip} x_q + \delta_{iq} x_p) L(\textbf{x}, t, t') + x_p x_q L_{,i}(\textbf{x}, t, t') \Big\}
    \end{aligned}
    \label{eq:D_spatial}
\end{equation}

In Eqs. (\ref{eq:L_spatial}) and (\ref{eq:D_spatial}), spatial Eshelby's tensor contains three functions with respect to the different orders of term ($t - t'$). Therefore, the time integrals for Eshelby's tensors can be determined with the following function $\mathcal{C}^{m, f}$ (m = 0, 1, 2): 

\begin{equation}
    \mathcal{C}^{m, f}(\textbf{x}, t) = \int_{t_{f-1}}^{t_f} [2 \alpha (t - t')]^m L(\textbf{x}, t, t') d t', \quad \text{where} \quad m = 0, 1, 2
    \label{eq:time_integral}
\end{equation}
where superscripts $m$ and $f$ of $\mathcal{C}^{m, f}$  represent the multiplication order of $2 \alpha (t - t')$ and time integral limits $t' \in [t_{f-1}, t_f]$, respectively. Substituting Eqs. (\ref{eq:L_spatial}) and (\ref{eq:D_spatial}) into (\ref{eq:Eshelby_tensor_time}) yields identical equations as Eq. (A.2-A.4) using symmetric properties in our recent work \cite{Wu_IJSS_2025}. Although it is tedious to copy these formulae, we reproduce these equations here for completeness and subsequent analysis:

\noindent (i) Time integrals for Eshelby's tensors $\overline{L}_{pq...}^f(\textbf{x}, t)$ for EHS:
\begin{equation}
    \begin{aligned}
    & \overline{L}^{f}(\textbf{x}, t) = \mathcal{C}^{0, f}(\textbf{x}, t)\\
    & \overline{L}^{f}_p(\textbf{x}, t) = \mathcal{C}^{1, f}_{,p}(\textbf{x}, t) + x_p \mathcal{C}^{0, f}(\textbf{x}, t) \\
    & \overline{L}^{f}_{pq}(\textbf{x}, t) = \mathcal{C}^{2, f}_{,pq}(\textbf{x}, t) + x_p \mathcal{C}^{2, f}_{,q}(\textbf{x}, t) + x_q \mathcal{C}^{2, f}_{,p}(\textbf{x}, t) +  (\delta_{pq} + 2 x_p x_q) \mathcal{C}^{1, f}_{,pq}(\textbf{x}, t) + x_p x_q \mathcal{C}^{0, f}(\textbf{x}, t)
    \end{aligned}
    \label{eq:L_final}
\end{equation}

\noindent (ii) Time integrals for Eshelby's tensors $\overline{D}_{ipq...}^f(\textbf{x}, t)$ for ETG:
\begin{equation}
    \begin{aligned}
    & \overline{D}_{i}^{f}(\textbf{x}, t) = -K \mathcal{C}_{,i}^{0, f}(\textbf{x}, t) \\ 
    & \overline{D}_{ip}^{f}(\textbf{x}, t) = -K \Big( \mathcal{C}_{,ip}^{1, f}(\textbf{x}, t)  + \delta_{ip} \mathcal{C}^{0, f}(\textbf{x}, t)  + x_p \mathcal{C}_{,i}^{0, f}(\textbf{x}, t) \Big) \\
    & \overline{D}_{ipq}^{f}(\textbf{x}, t) = -K \Big\{ \mathcal{C}_{,ipq}^{2, f}(\textbf{x}, t) + \delta_{ip} \mathcal{C}_{,q}^{2, f}(\textbf{x}, t) + \delta_{iq} \mathcal{C}_{,p}^{2, f}(\textbf{x}, t) + x_p \mathcal{C}_{,iq}^{2, f}(\textbf{x}, t) + x_q \mathcal{C}_{,ip}^{2, f}(\textbf{x}, t) \\ 
    & \qquad \qquad \quad + 2 (\delta_{ip} x_q + \delta_{iq} x_p) \mathcal{C}^{1, f}(\textbf{x}, t) + (\delta_{pq} + 2 x_p x_q) \mathcal{C}_{,i}^{1, f}(\textbf{x}, t) + (\delta_{ip} x_q + \delta_{iq} x_p) \mathcal{C}^{0, f}(\textbf{x}, t) +  x_p x_q \mathcal{C}^{1, f}_{,i}(\textbf{x}, t) \Big\}
    \end{aligned}
    \label{eq:D_final}
\end{equation}

The spatial and time Eshelby's tensors depend on the geometric shape of the inclusion. In the following, time Eshelby's tensors for polyhedral and ellipsoidal inclusions will be derived subsequently, and harmonic cases for polyhedral inclusions will be provided in \ref{appendix:harmonic}, which can be reduced to the 2D cases in the Supplemental Materials (Section 1). 

\section{Domain integrals over a polyhedral inclusion}
Rodin initiated Eshelby's tensors on an arbitrarily-shaped polygonal and polyhedral inclusion. Subsequently, Gao and Liu \cite{Gao2012} combined Eshelby's tensors with the strain-gradient theory, exploring size effects through characteristic length functions. The above two works paved the way for two recent works on polynomial-form Eshelby's tensors over polygonal \cite{Wu2021_jam_polygonal} and polyhedral \cite{Wu2021_polyhedral} inclusions, which are obtained through the 2D and 3D transformed coordinates. It should be noted these literature are only investigating elastostatic Eshelby's tensors, composed of the biharmonic and harmonic potentials, which are fundamentally different from Eshelby's tensors for transient/harmonic heat transfer. This section utilizes transformed coordinates in our recent works \cite{Wu2021_jam_polygonal, Wu2021_polyhedral} to frequency-/time-dependent loading and handles directional source terms in domain integrals. Because polynomial eigen-fields are considered, a series-form solution to transient heat transfer over the polygonal and polyhedral subdomains, respectively. As harmonic heat conduction can be considered a special case of general transient heat conduction, the following will elaborate on general transient heat conduction first and then reduce it to harmonic heat conduction. \textit{Note that domain integrals of polygonal inclusion are provided in Section 1 of the Supplemental Materials.}

\subsection{Derivation of spatial Eshelby's tensors}
\begin{figure}
\begin{subfigure}{.5\textwidth}
\centering
% include the first subplot
\includegraphics[width = 1\linewidth,height = \textheight,keepaspectratio]{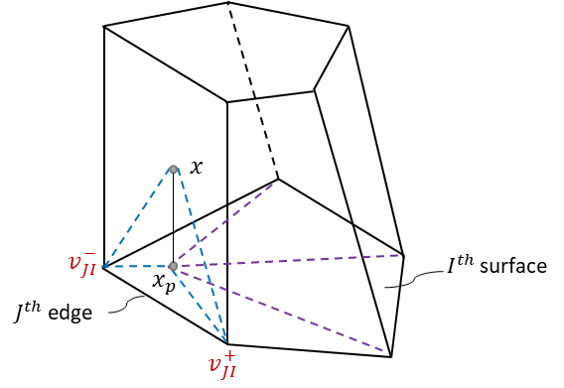}
\caption{}
\end{subfigure}
~
% include the second subplot
\begin{subfigure}{.5\textwidth}
\centering
\includegraphics[width = 0.7\linewidth,height = \textheight,keepaspectratio]{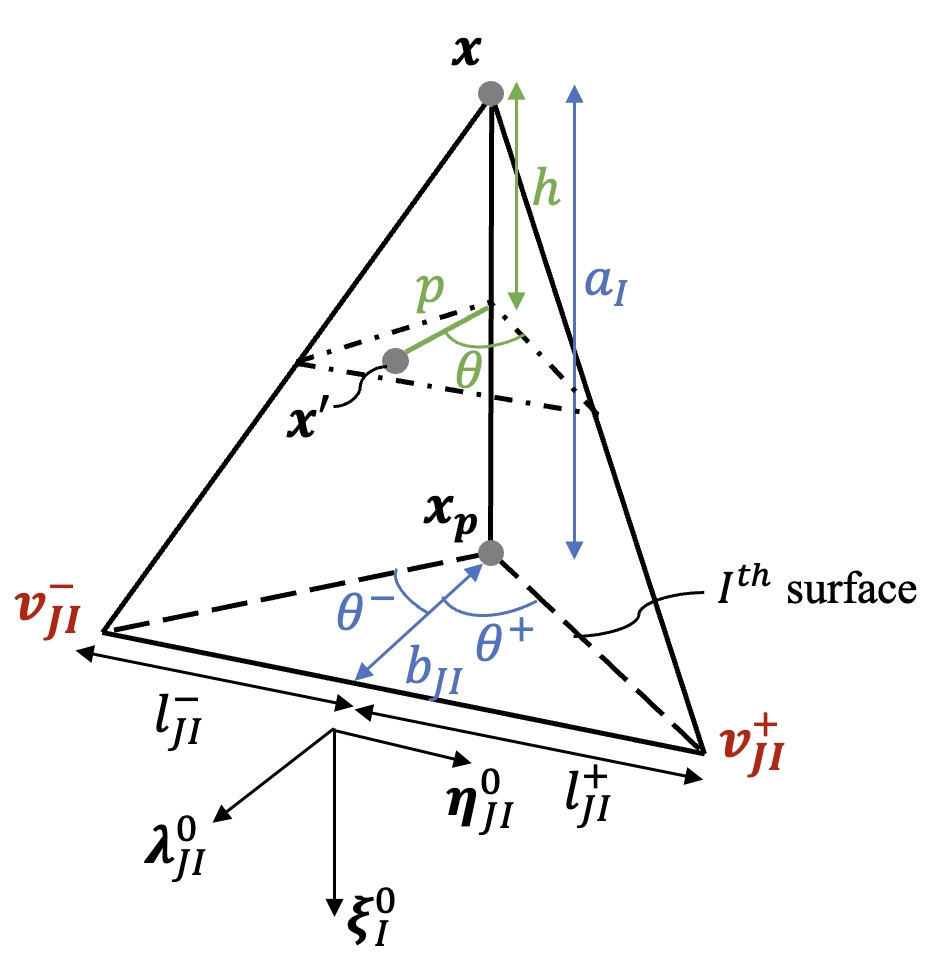}
\caption{}
\end{subfigure}
\caption{Schematic illustration of three-dimensional transformation coordinate (3DTC) on an arbitrary polyhedron, (a) a tetrahedral on the $I^{th}$ surface with $J^{th}$ edge; (b) dimensions in 3DTC of a tetrahedral based on field point $\textbf{x}$, its projection point $\textbf{x}_p$ and the $J^{th}$ edge of the $I^{th}$ surface}
\label{fig:3DTC}
\end{figure}

% \subsubsection{Three-dimensional transformed coordinate (3DTC)}
Consider an arbitrary polyhedron composed of $N_I$ surfaces, and each surface contains $N_{JI}$ edges embedded in an infinite domain $\mathcal{D}_{inf}$. In Fig. \ref{fig:3DTC} (a), a 3DTC \cite{Wu2021_polyhedral} is constructed by the field point ($\textbf{x}$), its projection point ($\textbf{x}_p$) on the $I^{th}$ surface, and the $J^{th}$ edge of the $I^{th}$ surface by two vertices $\bm{v}_{JI}^{\pm}$. Fig. \ref{fig:3DTC} (b) shows three base vectors (similar to the cylindrical coordinate) that, (i) $\bm{\xi}_{I}^{0}$ is the unit outward normal vector of the $I^{th}$ surface; and (ii) $\bm{\lambda}_{JI}^0$ and $\bm{\eta}_{JI}^0$ are unit normal, directional vectors of the $J^{th}$ edge in the surface, respectively. 

In the transformed coordinate of Fig. \ref{fig:3DTC} (b), $a_I$ denotes the distance from the field point to its projection surface, $b_{JI}$ the distance from the projection point to the $J^{th}$ edge, and $l_{JI}^{\pm} $ are the two distances between the foot of the perpendicular ($b_{JI}$) on the $J^{th}$ edge and each of the two vertices $\textbf{v}_{JI}^{\pm}$, which depends on $\textbf{x}$ as follows: 

\begin{equation}
    a_I = (x_i - (v_{JI}^+)_i) (\xi_{JI}^0)_i \quad \text{and} \quad b_{JI} = (x_i - (v_{JI}^+)_i) (\lambda_{JI}^0)_i \quad \text{and} \quad l_{JI}^{\pm} = (x_i - (v_{JI}^\pm)_i) (\eta_{JI}^0)_i
    \label{eq:TC_dim}
\end{equation}

The derivatives of a function $\mathcal{F}(a_I, b_{JI}, l_{JI})$ in the transformed coordinate can be evaluated through the chain rule, 

\begin{equation}
    \frac{\partial \mathcal{F}(a_I, b_{JI}, l_{JI})}{\partial x_i} = -(\xi_I^0)_i \frac{\partial \mathcal{F}}{\partial a_I} -(\lambda_{JI}^0)_i \frac{\partial \mathcal{F}}{\partial b_{JI}} -(\eta_{JI}^0)_i \frac{\partial \mathcal{F}}{\partial l_{JI}}
    \label{eq:3DTC_chain}
\end{equation}

% figures of 3DTC

%\subsubsection{Direct volume integrals}

The volume integrals over the polyhedron can be obtained through superposing volume integrals over the sub-tetrahedrons ($\Omega_{JI}$) along all surfaces. The volume integral in the 3DTC for any function $\mathcal{F}(|\textbf{x} - \textbf{x}'|)$ is written as follows: 

\begin{equation}
    \int_{\Omega} \mathcal{F}(|\textbf{x} - \textbf{x}'|) \thinspace d \textbf{x}' = \sum_{I=1}^{N_I} \sum_{J = 1}^{N_{JI}} \int_{\Omega_{JI}} \mathcal{F}(|\textbf{x} - \textbf{x}'|) \thinspace d \textbf{x}' = \sum_{I=1}^{N_I} \sum_{J = 1}^{N_{JI}} \int_{\theta^-}^{\theta^+} \int_{0}^{a_I} \int_{0}^{\frac{h b_{JI} \sqrt{1 + \tan^2 \theta}}{a_I}} \mathcal{F}(\sqrt{h^2 + p^2}) \thinspace p d p \thinspace d h \thinspace d \theta 
    \label{eq:integral_limits_3DTC}
\end{equation}
% define of integral limits:
where three integral variables $p, h, \theta$ and their integral limits are defined as Fig. \ref{fig:3DTC} (b), which is similar to a cylindrical coordinate; $p$ and $\theta$ represent polar and angle in the plane where $\textbf{x}'$ locates; $h$ defines the ``height'' of the plane. The integral limits can be defined as, $p \in [0, \frac{b_{JI} h}{a_I} \sqrt{1 + \tan^2 \theta}]$, $h \in [0, a_I]$ and $\theta \in [\theta^-, \theta^+]$, where $\theta^\pm = \tan^{-1} \left[ \frac{l_{JI}^\pm}{b_{JI}} \right]$. Hence, the distance between the field and source points is $|\textbf{x} - \textbf{x}'| = \sqrt{p^2 + h^2}$. 

Using the Taylor series expansion on the variable of $ \frac{-|\textbf{x} - \textbf{x}'|^2}{t - t'}$, the fundamental solution can be written as follows:

\begin{equation}
    G(\textbf{x}, \textbf{x}', t, t') = (4 \pi \alpha (t - t'))^{-\frac{3}{2}} \exp \left[ \frac{-|\textbf{x} - \textbf{x}'|^2}{4 \alpha (t - t')} \right] = (4 \pi \alpha (t - t'))^{-\frac{3}{2}}  \sum_{m=0}^{\infty} \frac{ (-1)^m }{m! (4 \alpha (t - t'))^{m}} |\textbf{x} - \textbf{x}'|^{2m}
    \label{eq:Taylor_fund}
\end{equation}

A straightforward evaluation of the volume integrals yields the spatial Eshelby's tensor as follows:

\begin{equation}
\begin{split}
    L(\textbf{x}, t, t') = & \frac{1}{(4 \pi (t - t'))^{3/2} K \sqrt{\alpha}} \sum_{m=0}^{\infty} \frac{(-1)^m}{m! (4 \alpha (t - t'))^m} \\ &  \sum_{I=1}^{N_I} \sum_{J=1}^{N_{JI}} \frac{a_I b_{JI}}{2(m + 1) (2m + 3)} \left[ A^m(a_I, b_{JI}, l_{JI}^+) - A^m(a_I, b_{JI}, l_{JI}^-) \right]
\end{split}
\label{eq:direct_vol}
\end{equation}
where
\begin{equation}
\begin{split}
    A^m(a_I, b_{JI}, l_{JI}) & = \frac{a_I^2}{b_{JI}^2} (a_I^2 + b_{JI}^2)^m \Big( l_{JI} F_1 \left[ \frac{1}{2}, -m, 1, \frac{3}{2}; \frac{-l_{JI}^2}{a_I^2 + b_{JI}^2}, \frac{-l_{JI}^2}{b_{JI}^2} \right] \Big) - \frac{(a_I)^{2m + 2}}{b_{JI}} \tan^{-1} \left[ \frac{l_{JI}}{b_{JI}} \right] \\ & + \frac{1}{a_I^2 + b_{JI}^2} \Big( l_{JI} (a_I^2 + b_{JI}^2 + l_{JI}^2 )^{1+m} \thinspace _2F_1\left[ 1, \frac{3}{2} + m; \frac{3}{2}; \frac{-l_{JI}^2}{a_I^2 + b_{JI}^2}\right] \Big)
\end{split}
\end{equation}
in which $_2F_1[a, b; c; x]$ is the hypergeometric function \cite{Bailey1935} as 
\begin{equation}
    _2F_1[a, b; c; x] = \frac{\Gamma(c)}{\Gamma(b) \Gamma(c-b)} \int_{0}^{1} \frac{t^{b-1} (1-t)^{c-b-1}}{(1 - t z)^a} \thinspace d t
\end{equation}
and $F_1[a, b_1, b_2, c; x, y]$ is Appell's series of the first kind \cite{Bailey1933} in the one-dimensional integral form as 
\begin{equation}
F_1[a, b_1, b_2, c; x, y] = \frac{\Gamma(c)}{\Gamma(a) \Gamma(c-a)} \int_{0}^{1} t^{a-1} (1 - t)^{c-a-1} (1 - x t)^{-b_1} (1 - y t)^{-b_2} \thinspace dt
\label{eq:Appell}
\end{equation}
where $\Gamma(s)$ is the complete gamma function. The weak singularity of $t^{-\frac{1}{2}}$ at $t \rightarrow 0$ can be handled by the singularity transformation \cite{Murota1982}. \textit{Numerical implementation has been provided in the supplemental source code ``Elementary\_functions.h'', including the weak singular and non-singular Appell's series.}

In Eq. (\ref{eq:super_disturb_time}), the evaluation of disturbed temperature requires partial differentiation of Green's function. For partial differentiation of domain integrals over a simple connected domain, it can be converted into surface (area) integrals using Gauss' theorem. Consider our target function, $G(\textbf{x}, \textbf{x}', t, t')$, the first partial derivative of its domain integral can be simplified with Gauss' theorem, 
\begin{equation}
\begin{split}
    \int_{\Omega} \frac{\partial}{\partial x_i} G(\textbf{x}, \textbf{x}', t, t') \thinspace d \textbf{x}' & = -\int_{\Omega} \frac{\partial}{\partial x_i'} G(\textbf{x}, \textbf{x}', t, t') \thinspace d \textbf{x}' = -\sum_{I=1}^{N_I} (\xi_{I}^0)_i \int_{\partial \Omega} \thinspace G(\textbf{x}, \textbf{x}', t, t') dA(\textbf{x}')
\end{split}
    \label{eq:Gauss}
\end{equation}
where the volume integral reduces to superpositions of surface integrals, which can be evaluated by direct area integral using the 3DTC. For example, Wu et al. \cite{Wu2021_polyhedral} used Gauss' theorem to derive the Eshelby's tensor for the static elastic problem. Alternatively, Stokes' theorem can be applied to convert surface integrals of the curl product as contour integrals. Gao and Liu \cite{Gao2012} has proposed that if a proper potential function can be found, the original surface integral can be further reduced into contour integrals, 

\begin{equation}
    \sum_{I=1}^{N_I} \int_{\partial \Omega} G(\textbf{x}, \textbf{x}') \thinspace dA(\textbf{x}') = \sum_{I=1}^{N_I} \sum_{J=1}^{N_{JI}} b_{JI} \int_{l_{JI}^-}^{l_{JI}^+} \frac{\mathcal{G}(a_I, b_{JI}, le, t, t')}{\sqrt{b_{JI}^2 + le^2}} \thinspace d le
    \label{eq:surface-contour}
\end{equation}
where the variable $le$ represents the integral point on the $J^{th}$ edge of the $I^{th}$ surface. Hence, when the source point is moving along the $J^{th}$ edge of the $I^{th}$ surface, the function has only one spatial variable $le$, as $a_I$ and $b_{JI}$ are fixed on that edge. Moreover, the potential function $\mathcal{G}(a_I, b_{JI}, le, t, t')$ and the target function $G(a_I, b_{JI}, le, t, t')$ are related by the following partial differential equation, 

\begin{equation}
G(a_I, b_{JI}, le, t, t') \sqrt{b_{JI}^2+le^2} = \mathcal{G}(a_I, b_{JI}, le, t, t') + b_{JI} \frac{\partial \mathcal{G}(a_I, b_{JI}, le, t, t')}{\partial b_{JI}} + le \frac{\partial \mathcal{G}(a_I, b_{JI}, le, t, t')}{\partial le}
    \label{eq:Gao_19_cpl}
\end{equation}
and 
\begin{equation}
    G(a_I, b_{JI}, le, t, t') = \frac{1}{(4 \alpha \pi (t - t'))^{\frac{3}{2}}} \exp \left[-\frac{a_I^2 + b_{JI}^2 + le^2}{4 \alpha (t - t')} \right]
\end{equation}
The solution to Eq. (\ref{eq:Gao_19_cpl}) is composed of the homogeneous and particular parts, which satisfies the condition $\mathcal{G}(a_I, b_{JI}, le, t, t') = 0$, when $b_{JI}$ and $le$ are zero \cite{Gao2012}. Using $\mathcal{G}(a_I, b_{JI}, le, t, t')$, the volume integral in Eq. (\ref{eq:Gauss}) can be written as contour integrals as follows: 

\begin{equation}
    \frac{\partial}{\partial x_i} \int_{\Omega} G(\textbf{x}, \textbf{x}', t, t') \thinspace d \textbf{x}' = -\sum_{I=1}^{N_I} (\xi_I^0)_i \int_{\partial \Omega} G(\textbf{x}, \textbf{x}', t, t') \thinspace d A(\textbf{x}') = -\sum_{I=1}^{N_I} (\xi_I^0)_i \sum_{J=1}^{N_{JI}} b_{JI} \int_{l_{JI}^-}^{l_{JI}^+} \frac{\mathcal{G}(a_I, b_{JI}, le, t, t')}{ \sqrt{b_{JI}^2 + le^2} } \thinspace d \thinspace le 
    \label{eq:Gao_17}
\end{equation}
where 
\begin{equation}
    \mathcal{G}(a_I, b_{JI}, le, t, t') = \frac{1}{4 \pi^{\frac{3}{2}} \sqrt{\alpha (t - t') (b_{JI}^2 + le^2)}} \left( \exp \left[ \frac{-a_I^2}{4 \alpha (t - t')} \right] - \exp \left[ \frac{-(a_I^2 + b_{JI}^2 + le^2)}{4 \alpha (t - t')} \right]  \right)
    \label{eq:mathcal_G}
\end{equation}

Use the series to represent the second exponential function of $\mathcal{G}(a_I, b_{JI}, le, t, t')$. The first-order partial derivative of the spatial Eshelby's tensor can be obtained, 

\begin{equation}
    \begin{split}
    L_{,i}(\textbf{x}, t, t') & = \sum_{I=0}^{N_I} (-\xi_I^0)_i \sum_{J=0}^{N_{JI}} 
    \Bigg\{ \frac{\alpha}{4 \pi^{\frac{3}{2}} K (\alpha (t - t'))^{\frac{1}{2}}} \exp \left[ \frac{-a_I^2}{4 \alpha (t - t')} \right] \Big( \tan^{-1} \left[ \frac{l_f^+}{b_{JI}} \right] - \tan^{-1} \left[ \frac{l_f^-}{b_{JI}} \right] \Big) \\ 
    & - \sum_{m = 0}^\infty \frac{\alpha}{2 \pi^{\frac{3}{2}} K (4 \alpha (t - t'))^{m + \frac{1}{2}}} \Big( 
      B^m(a_I, b_{JI}, l_{JI}^+) - B^m(a_I, b_{JI}, l_{JI}^-) \Big) \Bigg\}
    \end{split}
    \label{eq:stokes_spatial}
\end{equation}
where
\begin{equation}
    \begin{split}
    B^m (a_I, b_{JI}, le) = \frac{(-1)^m}{m!} \frac{le}{b_{JI}} (a_I^2 + b_{JI}^2)^m F_1 \left[ \frac{1}{2}, -m, 1, \frac{3}{2}; \frac{-le^2}{a_I^2 + b_{JI}^2}, \frac{-le^2}{b_{JI}^2} \right]
    \end{split}
\end{equation}
in which $F_1$ is defined in Eq. (\ref{eq:Appell}).

\subsection{Derivation of the time Eshelby's tensors}
Spatial Eshelby's tensors by direct volume integral or Stokes' theorem are provided in Eqs. (\ref{eq:direct_vol}) and (\ref{eq:stokes_spatial}), respectively. The time Eshelby's tensors are constructed by integrating the spatial Eshelby's tensors over the time interval $t' \in [t_{f-1}, t_f]$, as expressed in Eq. (\ref{eq:time_integral}). Applying this time integration to Eqs. (\ref{eq:direct_vol}) and (\ref{eq:stokes_spatial}) yield the following results: 

\begin{equation}
    \begin{split}
    & \mathcal{C}^{n, f}(\textbf{x}, t) = \sum_{m = 0}^{\infty} \frac{(2 \alpha)^n (-1)^m}{2 (2 m + 1 - 2 n) K (4 \alpha)^{m + \frac{1}{2}} \pi^{\frac{3}{2}} m!} \Big( (t - t_f)^{\frac{1}{2} + n -m} - (t - t_{f-1})^{\frac{1}{2} + n -m} \Big) \\ %
    & \quad \quad \quad \quad \quad \times  \sum_{I=1}^{N_I} \sum_{J = 1}^{N_{JI}} \frac{a_I b_{JI}}{2 (m + 1) (2 m + 3)} 
    \Big( A^m(a_I, b_{JI}, l_{JI}^+) - A^m(a_I, b_{JI}, l_{JI}^-) \Big)
    \end{split}
    \label{eq:3D_time}
\end{equation}
and
\begin{equation}
    \begin{split}
     \mathcal{C}_{,i}^{n, f}(\textbf{x}, t) & = \sum_{I=0}^{N_I} (\xi_I^0)_i \sum_{J=0}^{N_{JI}} \Bigg\{ \frac{(a_I^2)^{\frac{1}{2} + n}}{2^{3 + n} K \pi^{\frac{3}{2}}} \Big( \Gamma \left[ -\frac{1}{2} - n, \frac{a_I^2}{4 \alpha (t - t_f)} \right] - \Gamma \left[ -\frac{1}{2} - n, \frac{a_I^2}{4 \alpha (t - t_i)} \right] \Big) \\ & \Big( \tan^{-1} \left[ \frac{l_f^+}{b_{JI}} \right] - \tan^{-1} \left[ \frac{l_f^-}{b_{JI}} \right] \Big)  + \sum_{m = 0}^\infty \frac{2^{n - 2m - 1} \alpha^{n - m + \frac{1}{2}}}{(2 m - 2n - 1)\pi^{\frac{3}{2}} K} \Big( (t - t_f)^{\frac{1}{2} + n -m} - (t - t_{f-1})^{\frac{1}{2} + n -m} \Big) \\ 
    & \Big( 
      B^m(a_I, b_{JI}, l_{JI}^+) - B^m(a_I, b_{JI}, l_{JI}^-) \Big) \Bigg\}
    \end{split}
    \label{eq:stokes_time}
\end{equation}
respectively, where $n = 0, 1, 2$. The substitution of Eqs. (\ref{eq:3D_time}) and (\ref{eq:stokes_time}) into Eqs. (\ref{eq:L_final}) and (\ref{eq:D_final}) provides the time Eshelby's tensors. It should be noted that readers may assume time-varying quantities, taking the form of a linear function $ a + b t$. Because the spatial integrals has been evaluated and the function $B^m$ is time-independent, the extension from Eq. (\ref{eq:stokes_spatial}) is straightforward by multiplying the linear time variable and then integrate.

It should be emphasized that when $t = t_{f}$ and $n < m$, evaluating evaluating $(t - t_f)^{n-m}$ exists weak mathematical singular issues, which can be handled through interchanging integral sequences as elaborated in \ref{appendix:sing}. The treatment of interchanging integral sequence, first time then spatial, is a well-established technique in boundary element method \cite{Gupta1995}, which successfully separate the weakly singular kernel function $\frac{1}{|\textbf{x} - \textbf{x}'|}$ in the transient analysis. \ref{appendix:sing} rigorously proved that when $n = 0, 1, 2$, the separated (time-independent) kernel functions are, $\frac{1}{|\textbf{x} - \textbf{x}'|}, |\textbf{x} - \textbf{x}'|$, and $|\textbf{x} - \textbf{x}'|^{3}$, respectively. Moreover, the convergence of the sum of the series is elaborated in \ref{appendix:con}.

\section{Domain integrals over an ellipsoidal inclusion}
Consider an ellipsoidal subdomain $\Omega$ embedded in the infinite matrix, and three semi-axes of the ellipsoid are $a_1, a_2$ and $a_3$, respectively. Without the loss of any generality, let $a_1, a_2$ and $a_3$ be parallel to $x_1, x_2$ and $x_3$ in the Cartesian coordinate, and the center of the ellipsoid $\textbf{x}^C = \textbf{0}$. The ellipsoidal subdomain can be described through a quadratic equation, 

\begin{equation}
    \frac{x_1^2}{a_1^2} + \frac{x_2^2}{a_2^2} + \frac{x_3^2}{a_3^2} \leq 1
    \label{eq:ellip_gov}
\end{equation}

Michelitsch and colleagues first proposed to use the convolution property of the Fourier space to evaluate Helmholtz's\cite{Michelitsch2003} or retarded \cite{Wang2005} potentials. The integral strategy exhibits two main advantages over the classic way (\cite{Mura1987}): (a) the mathematical singularity for interior field points is well handled and (b) integral limits for interior and exterior points have clear definition. Therefore, domain integrals of transient heat conduction for ellipsoidal inclusion will be derived in the Fourier space. Note that for harmonic heat transfer, the fundamental solution is a revised Helmholtz's potential. Helmholtz's potential with polynomial-form source density has been provided by our recent work \cite{Wu2024-prsa}. 

\subsection{Target and characteristic functions in the Fourier space}
Eqs. (\ref{eq:L_spatial}) and (\ref{eq:D_spatial}) have indicated that spatial Eshelby's tensor for linear or quadratic eigen-fields can be derived by the spatial Eshelby's tensor for the uniform eigen-fields. Therefore, the following target function is the transient fundamental solution. Without the loss of any generality, let $t' = 0$ and conduct the Fourier transform of the transient fundamental solution, the target function, in Eq. (\ref{eq:fund_time}) as follows: 

\begin{equation}
    \tilde{G}(\textbf{k}, t) = \int_{0}^{\pi} \int_{0}^{2 \pi} \int_0^\infty (4 \pi \alpha t )^{\frac{-3}{2}} \exp \left[ \frac{-x^2}{4 \alpha t} - i \textbf{k} \cdot \textbf{x} \right] H(t) \thinspace x^2 d x \thinspace d \theta \thinspace \sin \gamma \thinspace d \gamma = \exp \left[ - \alpha k^2 t \right] H(t)
    \label{eq:Fourier_fund}
\end{equation}
where $\textbf{k}$ represents vectors in the Fourier space, and $k = \sqrt{k_i k_i}$; dot products of vectors $\textbf{k} \cdot \textbf{x} = k x \cos \gamma$, and $\gamma$ stands for the angle between the directional normal vectors of \textbf{k} and \textbf{x}. Since the eigen-fields only exist within the subdomain, quantities within the ellipsoidal subdomain can be written as a characteristic function \cite{Michelitsch2003}. Using a mapping spherical coordinate, 

\begin{equation}
    \Theta(1 - \zeta) = \begin{cases} 1 & \zeta \in [0, 1) \\ 0 & \zeta \in (1, +\infty) \end{cases}, \quad \zeta^2 = \frac{x_i x_i}{a_I^2} = \frac{x_1^2}{a_1^2} + \frac{x_2^2}{a_2^2} + \frac{x_3^2}{a_3^2}
    \label{eq:charac}
\end{equation}
where the capital index $I$ does not trigger dummy index summation of $i$, but changes its value during the summation following Mura's index notation \cite{Mura1987}. The Fourier transform of the characteristic function is written as: 

\begin{equation}
\begin{split}
    \tilde{\Theta}(\textbf{k}) & = a_1 a_2 a_3 \int_0^\pi \int_0^{2 \pi} \int_{0}^{\infty} \Theta(1 - \zeta) \exp[-i \textbf{x} \cdot \textbf{k}] \zeta^2 d\zeta \thinspace d \theta \thinspace \sin \gamma d \gamma \\
    & = 2 \pi a_1 a_2 a_3 \int_0^\pi \int_{0}^{1} \exp[-i \zeta k S \cos \gamma] \zeta^2 d\zeta \thinspace \sin \gamma d \gamma = \frac{4 \pi a_1 a_2 a_3}{k^3 S^3} \left( \sin k S - k S \cos k S \right)
\end{split}
\label{eq:Charc_fourier}
\end{equation}
where $\textbf{x} \cdot \textbf{k} = (\zeta a_I n_i) (k \hat{k}_j) \delta_{ij} = p k S \cos \gamma$; $S = (a_I^2 \hat{k}_i \hat{k}_i)^{1/2}$ and $\hat{k}_i = k_i / k$ is the unit vector in the Fourier space. Thanks to the convolution property in the Fourier space, the domain integral of the transient fundamental solution in the Fourier space is the product of Eqs. (\ref{eq:Fourier_fund}) and (\ref{eq:Charc_fourier}) as follows: 

\begin{equation}
    \tilde{L}(\textbf{k}, t) = \exp \left[ -\alpha k^2 t \right] \frac{4 \pi a_1 a_2 a_3}{C_p k^3 S^3} \left( \sin k S - k S \cos k S \right) H(t)
    \label{eq:dm_fourier}
\end{equation}
The following will conduct the inverse Fourier transform to obtain explicit formulae in the spherical coordinate. 
\subsection{Inverse Fourier transform}
To conduct the inverse Fourier transform, another coordinate mapping that the vector in the Fourier space is applied, $k_i = p \frac{e_i}{a_I}$, where $\textbf{e}$ is a unit vector. Based on such notation, previous quantities ($k, \hat{k}_i, S$) can be expressed, 

\begin{equation}
    k = p \sqrt{\frac{e_n e_n}{a_N^2} }, \quad \hat{k}_i = \frac{e_i}{a_I \sqrt{\frac{e_n e_n}{a_N^2} }},\quad S = \sqrt{\frac{a_I^2 e_i e_i}{a_I^2 {\frac{e_n e_n}{a_N^2} }}} = \frac{1}{\sqrt{\frac{e_n e_n}{a_N^2} }}= \frac{p}{k}
    \label{eq:express_KS}
\end{equation}
where  we can also write $\hat{k}_i = S e_i / a_I$. Using the coordinate notation $k_i = p \frac{e_i}{a_I}$, we conduct the inverse Fourier transform of $\tilde{L}(\textbf{k}, t)$ in Eq. (\ref{eq:dm_fourier}) as follows: 

\begin{equation}
\begin{split}
     L(\textbf{x}, t)  = & (2 \pi)^{-3} \frac{4 \pi a_1 a_2 a_3}{C_p} \int_{|\hat{\textbf{k}}| = 1} \int_0^\infty \exp[i \textbf{k} \cdot \textbf{x} - \alpha k^2 t] \frac{\sin k S - k S \cos k S}{k^3 S^3} \thinspace k^2 \thinspace d k \thinspace d A(\hat{\textbf{k}}) H(t) \\ 
    = & \frac{H(t)}{2 \pi^2 C_p} \int_{|\textbf{e}| = 1} \int_0^\infty \exp \left[ i p \mathcal{B} - \frac{p^2 \alpha t}{S^2} \right] \frac{\sin p - p \cos p}{p S^2} \thinspace d p \thinspace d A(\textbf{e}) \\
    = & \frac{H(t)}{8 \pi C_p} \int_{|\textbf{e}| = 1} \left\{  \Big( \text{Erf} \left[ \frac{(1- \mathcal{B}) S}{\sqrt{4 \alpha t}} \right] + \text{Erf} \left[ \frac{(1 + \mathcal{B}) S}{\sqrt{4 \alpha t}} \right] \Big) \right. \\ 
    & \left. + \frac{S}{\sqrt{\pi \alpha t}} \Big( \exp \left[ \frac{-S^2}{4 \alpha t} (1 + \mathcal{B})^2 \right] + \exp \left[ \frac{-S^2}{4 \alpha t} (1 - \mathcal{B})^2 \right] \Big) \right\} d A(\textbf{e})
\end{split}
    \label{eq:L_inverse0}
\end{equation}
%%%
where $\textbf{k} \cdot \textbf{x} = (\zeta a_I n_i) (p \frac{e_j}{a_J}) \delta_{ij} = p \mathcal{B}$ and $\mathcal{B} = \zeta n_i e_i$; and Erf[.] is the error function. Due to the singularity of the Green's function, it is significant to discuss whether the field points are interior or exterior of the integral domain. For Helmholtz's potential, Michelitsch et al. \cite{Michelitsch2003} separate two branches of solutions based on the parameter $\mathcal{B}$: (a) when the field point is interior, $\mathcal{B} = \zeta e_i n_i$ is bounded as $\mathcal{B} \in (-1, 1)$ because $\zeta \in [0, 1)$ and $e_i n_i \in [-1, 1]$; (b) when the field point is exterior, $\mathcal{B} \in (-\infty, +\infty)$. The necessity for such a discussion lies in that expression after radial inverse Fourier transform is a conditional expression, i.e., some parts rely on $\mathcal{B} \in [1, +\infty)$, $\mathcal{B} \in (-1, 1)$ or $\mathcal{B} \in (-\infty, -1]$, as seen in Helmholtz's potential for the three cases and Newtonian potential for only $\mathcal{B} \in (-1, 1)$. Therefore, for an interior point, the integral limits of the mapping spherical shell is a complete one; whereas for an exterior point, the auxiliary discussion must be made on integral limits. 

However, unlike any previous potential functions, domain integrals of the transient fundamental solution exhibit an attractive feature that it does not need any discussion on conditional expressions, which implies that the integral limits for interior and exterior field points are the same as a complete spherical shell. The interesting finding is confirmed by Fig. \ref{fig:Contour_cuboid} on a cuboid inclusion. Therefore, Eq. (\ref{eq:L_inverse0}) can be rewritten as a complete spherical shell integral, 

\begin{equation}
\begin{split}
    L(\textbf{x}, t) & = \frac{H(t)}{8 C_p} \int_{0}^{\pi} \int_0^{2 \pi} \left\{ \Big( \text{Erf} \left[ \frac{(1- \mathcal{B}) S}{\sqrt{4 \alpha t}} \right] + \text{Erf} \left[ \frac{(1 + \mathcal{B}) S}{\sqrt{4 \alpha t}} \right] \Big) \right. \\ & \left. + \frac{S}{\sqrt{\pi \alpha t} C_p} \Big( \exp \left[ \frac{-S^2}{4 \alpha t} (1 + \mathcal{B})^2 \right] + \exp \left[ \frac{-S^2}{4 \alpha t} (1 - \mathcal{B})^2 \right] \Big) \right\} d \theta \thinspace \sin \gamma d \gamma
\end{split}
    \label{eq:L_inverse1}
\end{equation}

\subsection{Special case: sphere}
For a spheroid, Eq. (\ref{eq:L_inverse1}) can be further simplified as a one-dimensional integral \cite{Wu2024-prsa}, and explicit results can be derived for the spherical subdomain, where $S$ equals the radius $a$. Conduct the spherical shell integral in Eq. (\ref{eq:L_inverse1}) as follows:

% I placed this content to increase readbility, to avoid too large gaps in derivation. 

\begin{equation}
    \begin{split}
    L(\textbf{x}, t) & = \frac{H(t)}{8 C_p} \int_{0}^{\pi} \int_0^{2 \pi} \Big( \text{Erf} \left[ \frac{(1- \zeta \cos \gamma) a}{\sqrt{4 \alpha t}} \right] + \text{Erf} \left[ \frac{(1 + \zeta \cos \gamma) a}{\sqrt{4 \alpha t}} \right] \Big) \\ & + \frac{a}{\sqrt{\pi \alpha t} C_p} \Big( \exp \left[ \frac{-a^2}{4 \alpha t} (1 + \zeta \cos \gamma)^2 \right] + \exp \left[ \frac{-a^2}{4 \alpha t} (1 - \zeta \cos \gamma)^2 \right] \Big) d \theta \thinspace \sin \gamma d \gamma \\ 
    & = \frac{H(t) \alpha}{2 K} \Big( \text{Erf} \left[ \frac{a - x}{\sqrt{4 \alpha t}} \right] + \text{Erf} \left[ \frac{a + x}{\sqrt{4 \alpha t}} \right] \Big) + \frac{ 
\alpha^{\frac{3}{2}} \sqrt{t} H(t)}{\sqrt{\pi} K x} \Big( \exp  \left[-\frac{(a + x)^2}{4 \alpha t} \right] - \exp  \left[-\frac{(a - x)^2}{4 \alpha t} \right] \Big)
\end{split}
    \label{eq:L_sphere}
\end{equation}
where $\mathcal{B} = \zeta \cos \gamma$ and $\zeta = \frac{x}{a}$ are used.

The transition functions $\mathcal{C}^{n, f}(\textbf{x}, t)$ for time Eshelby's tensors can be obtained by time integral as follows, 

\begin{equation}
    \mathcal{C}^{n, f}(\textbf{x}, t) = \mathcal{E}^{n}(\textbf{x}, t - t_{f}) - \mathcal{E}^{n}(\textbf{x}, t - t_{f-1})
\end{equation}
where 
\begin{equation}
    \begin{split}
     \mathcal{E}^0(\textbf{x}, t) &= \frac{1}{12 K x} \Big\{ 
    \sqrt{\frac{4 \alpha t }{\pi}} \left( -(2 a - x) (a + x) + 4 \alpha t  \right) \exp \left[ \frac{-(a + x)^2}{4 \alpha t} \right] \\ & + \left( (2a + x) (a - x) - 4 \alpha t \right) \exp \left[ \frac{-(a - x)^2}{4 \alpha t}  \right]  
    + \left( (2a + x) (a - x)^2 + 6 x \alpha t \right) \text{Erf}\left[ \frac{a - x}{\sqrt{4 \alpha t}} \right] \\ & 
    + \left( -(2a - x) (a + x)^2 + 6 x \alpha t \right)  \text{Erf} \left[ \frac{a + x}{\sqrt{4 \alpha t}} \right]
    \Big\}
    \end{split}
    \label{eq:sphere_time_uni}
\end{equation}
and 
\begin{equation}
    \begin{split}
    \mathcal{E}^1(\textbf{x}, t) & = \frac{1}{120 x K} \Big\{ \sqrt{\frac{4 \alpha t}{\pi}} \left( (4a - x) (a + x)^3 - 2 (4a - x) (a + x) \alpha t + 48 \alpha^2 t^2 \right) \exp [\frac{-(a + x)^2}{4 \alpha t}] \\ & 
    \left( -(4a + x) (a - x)^3 + 2 (4a + x) (a - x) \alpha t - 48 \alpha^2 t^2 \right) \exp [\frac{-(a - x)^2}{4 \alpha t}] \\ & 
    - \left( (4a + x) (a-x)^4 - 60 x \alpha^2 t^2 \right) \text{Erf} \left[ \frac{a - x}{\sqrt{4 \alpha t}} \right] + \left( (4a - x) (a + x)^4 + 60 x \alpha^2 t^2 \right) \text{Erf} \left[ \frac{a + x}{\sqrt{4 \alpha t}} \right]
    \Big\}
    \end{split}
    \label{eq:sphere_time_lin}
\end{equation}
and 
\begin{small}
\begin{equation}
    \begin{split}
    \mathcal{E}^2(\textbf{x}, t) & = \frac{1}{1260 x K} \Big\{ \sqrt{\frac{4 \alpha t}{\pi}} \left( -(6a - x) (a+x)^5 + 2 (6a - x) (a + x)^3 \alpha t - 12 (6a - x) (a + x) \alpha^2 t^2 + 720 \alpha^3 t^3 \right) \\ 
    & \exp \left[ \frac{-(a + x)^2}{4 \alpha t} \right] + \left( (6a + x) (a - x)^5 - 2 (6a + x) (a-x)^3 \alpha t + 12 (6a + x) (a - x) \alpha^2 t^2 - 720 \alpha^3 t^3 \right) \\
    & \exp \left[ \frac{-(a-x)^2}{4 \alpha t} \right]  + \left( (6a + x) (a - x)^6 + 840 x \alpha^3 t^3 \right) \text{Erf}\left[ \frac{a - x}{\sqrt{4 \alpha t}} \right] - \left( (6a - x) (a + x)^6 - 840 x \alpha^3 t^3 \right) \text{Erf}\left[ \frac{a + x}{\sqrt{4 \alpha t}} \right]
    \Big\}
    \end{split}
    \label{eq:sphere_time_qua}
\end{equation}
\end{small}
where Eq. (\ref{eq:sphere_time_uni}), Eq. (\ref{eq:sphere_time_lin}), and Eq. (\ref{eq:sphere_time_qua}) recovers the integral results in \cite{Wu_IJSS_2025} using transformed spherical coordinates. When $t \rightarrow 0$ and $(a \pm x) / \sqrt{4 \alpha t} \rightarrow \infty$, the error function becomes $1$. Therefore, no special treatment is required in the case of spherical inclusions. Note that formulae of cylindrical circular inclusions can be obtained by setting the third axis as infinity, which reduces to one-dimensional integrals of two Bessel's function of the first kind. \textit{The numerical implementation of spherical Eshelby's tensors is provided in Section 3 of the Supplemental Materials.}

\section{Numerical case studies}
The method of transformed coordinates for arbitrarily shaped inclusions has been verified through comparison with analytical solutions in literature, such as elastostatic Eshelby's tensors of ellipsoidal inclusions \cite{Gao2012, Wu2021_polyhedral}. The method of Fourier transform has been verified through comparison with spherical solutions in \cite{Michelitsch2003}, such as Helmholtz's potential first proposed by \cite{Mikata1990}. Note that our recent work \cite{Wu2024-prsa} have explicitly shown that the method of Fourier transform can recover the celebrated biharmonic/harmonic potentials over spherical/spheroidal inclusions, which is elaborated in Section 4 of the supplemental information. As discussed above, it is a traditional manner in the literature to first compare new proposed solutions with some closed-form results, and then apply it explore more features of other geometric shapes. 

To verify the formulae presented in Section 3 and Section 4, this section utilizes three steps: (\textit{The guidance on the source code is provided in Section 2 of the Supplemental Materials.})

\begin{enumerate}
   \item Polyhedron-approximated spherical inclusion: implement Eqs. (\ref{eq:Helmholtz_direct}) and (\ref{eq:Helmholtz_stokes}), and compare results provided by closed-form Helmholtz's potential (only available for spheres) \cite{Mikata1990}. 
    
    \item Polyhedron-approximated spherical inclusion: implement Eqs. (\ref{eq:direct_vol}) and (\ref{eq:stokes_spatial}), and compare results provided by closed-form solution in Eq. (\ref{eq:L_sphere}) (only available for spheres).

    \item Cuboid inclusion: implement Eqs. (\ref{eq:direct_vol}) and (\ref{eq:stokes_spatial}), and compare with results provided by the method of Fourier transform, which is only available as numerical presentations.
\end{enumerate}

Step (i) ensures the transformed coordinates can be applied for transient/harmonic heat transfer, and Step (ii) ensures the transformed coordinate can reproduce the closed-form solution proposed by the method of Fourier transform. Since closed-form solutions are independent of any approximations, it is rational to select them as references. After verifications of both integral schemes, it is important to demonstrate their applicability on cuboid inclusions in Step (iii)  due to three reasons: (1) unlike spheres/ellipsoids, the cuboid is not approximated; (2) its characteristic function is straightforward; (3) the cuboid inclusions have potential application solving arbitrarily-shaped inhomogeneities with the fast-Fourier transform algorithm.

\subsection{Verification of domain integrals}
The formulation has been implemented in the computational program for actual case studies and applications. Domain integrals of a spherical inclusion can not only be presented as closed-form expressions in the previous section, but also be used for verification of domain integrals of polyhedrons as a sphere can be approximated by many polyhedrons \cite{Wu2021_polyhedral}. In the following, the domain integral of a sphere will be verified first by a recent solution on Helmholtz's potential, and then by domain integrals of the transient fundamental solution on the corresponding multiple polyhedrons. The material properties and dimension are specified as: (a) the spherical inclusion with radius $a = 0.1$ m is located at the origin; (b) the thermal conductivity $K = 0.05 W / (m \cdot K)$, the heat transfer coefficient $\alpha = 0.05 m^2 / s$, and the complex parameter (for harmonic thermal load) $\beta = 10 (1 + i) m^{-1}$. The series-form solution for polyhedral inclusion requires an approximated sphere, which is approximated by $N_I = 229, 683, 3,664$ and $18,549$-surface polyhedrons. In general, an approximated sphere with more surfaces will improve the accuracy. 

\subsubsection{Helmholtz's potential function over a spherical inclusion}
\begin{figure}
    \centering
    \includegraphics[width = 1\textwidth, keepaspectratio]{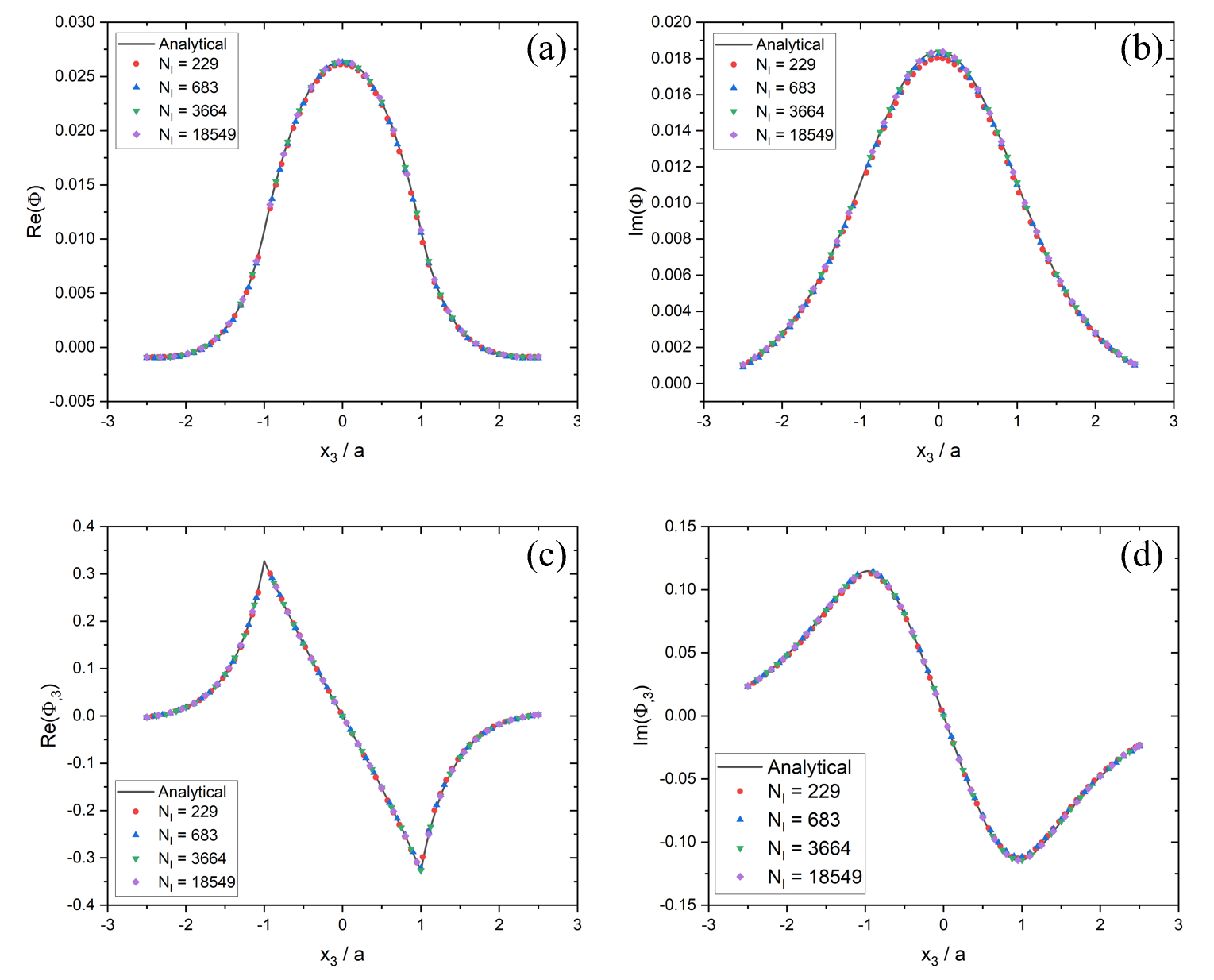}
    \caption{Verification of domain integrals of Helmholtz's potential over an approximated sphere by $N_I = 229$, $683$, $3,664$, and $18,549$-surface polyhedrons, when $x_3 \in [-2.5, 2.5]$ a. (a), (b) are real and imaginary parts of $\Phi$ by direct volume integral; (c), (d) are real and imaginary parts of $\Phi_{,3}$ by Stokes' theorem}
    \label{fig:helmholtz}
\end{figure}

Figs. \ref{fig:helmholtz} (a-d) compare domain integrals of Helmholtz's potential over an approximated sphere by $N_I = 229$, $683$, $3,664$, and $18,549$ surface polyhedrons, where the number of series is $20$. The analytical formulae was first proposed by Mikata and Nemat-Nasser \cite{Mikata1990} and recovered by Michelitsch et al. \cite{Michelitsch2003}. Results in Figs. \ref{fig:helmholtz} (a-b) were evaluated by the direct volume integral method as Eq. (\ref{eq:Helmholtz_direct}). For the real part, all approximated can provide good accuracy compared with the analytical solution, while some discrepancies between the case ``$N_I = 229$'' and other curves in Fig. \ref{fig:helmholtz} (b) can be found. The relative error is generally caused by approximation, and the error can be reduced by introducing more surfaces. Subsequently, Figs. \ref{fig:helmholtz} (c-d) compare real and imaginary parts of $\Phi_{,3}$ evaluated by Stokes' theorem as Eq. (\ref{eq:Helmholtz_stokes}). Note that the domain integrals in the  two methods require evaluation integrals $N_I \times N_{JI}$ times, but the application of Stokes' theorem simplifies the calculation. For real and imaginary parts, all approximated solutions provide good agreement with the analytical solution when $x_3 / a \in [-2.5, 2.5]$. Although $Re(\Phi_{,3})$ exhibits a sharp corner on the interface of the inclusion, its imaginary part exhibits a smooth variation. Note that the accuracy of series-form solution decreases with the distance from the inclusion, and such effects are observed and clarified in \ref{appendix:con}. 

\subsubsection{Domain integrals of transient Green's function over a spherical inclusion}
The previous subsection verifies two integral schemes by the classic Helmholtz's potential over a spherical inclusion, and this subsection aims to compare closed-form solution as Eq. (\ref{eq:L_sphere}) and series-form solution as Eqs. (\ref{eq:direct_vol}) and (\ref{eq:stokes_spatial}). Since the domain integrals of the uniform, linear and quadratic terms share similar functions, this subsection only verifies the uniform one. For the completeness of the verification, the linear and quadratic time Eshelby's tensors are verified in \ref{appendix:eshel}. According to \ref{appendix:con}, Fig. \ref{fig:convergence} (a) indicated the number of series ($n_{max} = 10$) is a rational choice with acceptable accuracy for the series-form solution when $x_3 / a \in [-2.5, 2.5]$. 

\begin{figure}
    \centering
    \includegraphics[width = 1\linewidth,height = \textheight,keepaspectratio]{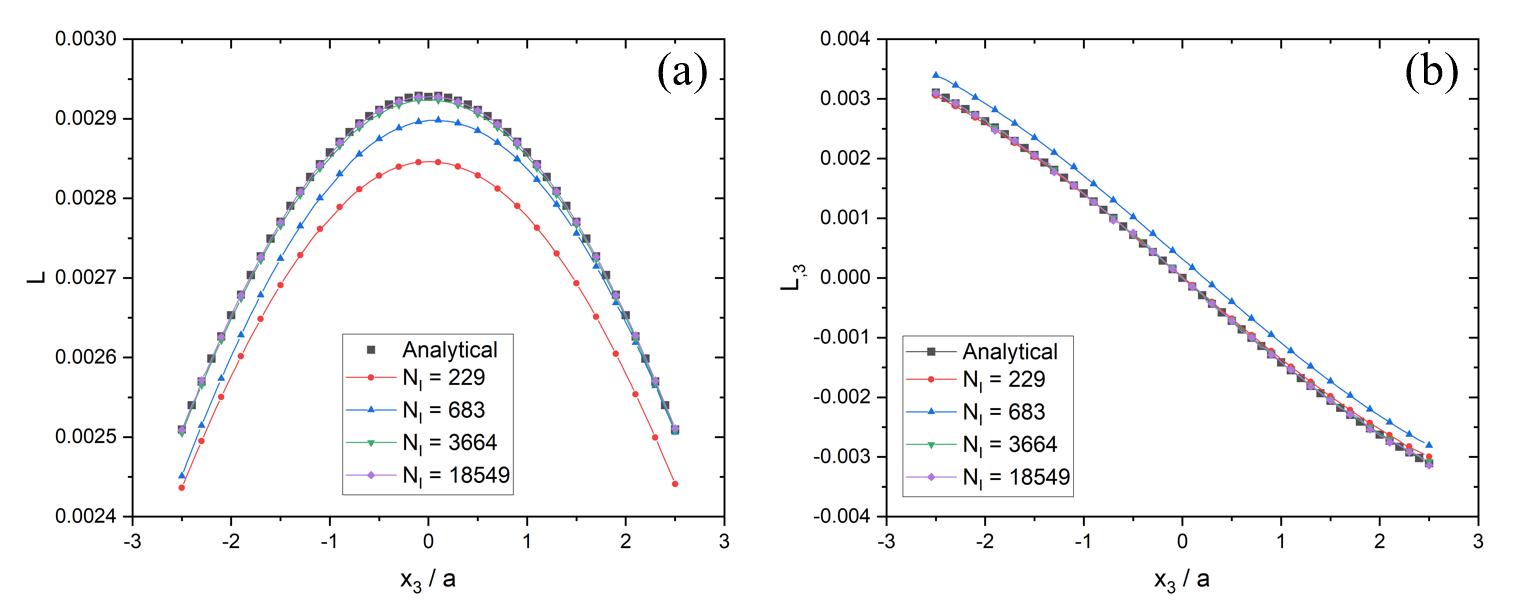}
    \caption{Reproduction and verification of the analytical spherical integral with radius $a = 0.1$ m, $t = 2 s$ , $\alpha = 0.05 m^2 / s$ along the vertical center line 
    $x_3 \in [-2.5, 2.5] a$ on (a) different $N_I$-surface polyhedron ($229$, $683$, $3,664$, and $18,549$) using direct volume integral scheme as Eq. (\ref{eq:direct_vol}); (b) different $N_I$-surface polyhedron using Stokes' theorem integral scheme as Eq. (\ref{eq:stokes_spatial}). ($n_{max} = 10$)}
    \label{fig:verify_spher_transient}
\end{figure}

Figs. \ref{fig:verify_spher_transient}(a) and  \ref{fig:verify_spher_transient}(b) compares the domain integral and its partial derivative (w.r.t $x_3$) using the direct volume integral and Stokes' theorem integral scheme, respectively. It can be observed that when the sphere is approximated with more polyhedrons, approximated results become closer to the analytical solution as Eq. (\ref{eq:L_sphere}). Comparing $N_I = 229$ with the analytical solution in both Figs. \ref{fig:verify_spher_transient}(a) and \ref{fig:convergence}(b), more significant discrepancies can be observed. The phenomenon can be interpreted as two main errors: (i) minor variation of the domain integral and (ii) accuracy in evaluating the Gauss hypergeometric function and Appell's series. 

% two figures on Integral and error comparison

\begin{figure}
    \centering
    \includegraphics[width = 1\linewidth,height = \textheight,keepaspectratio]{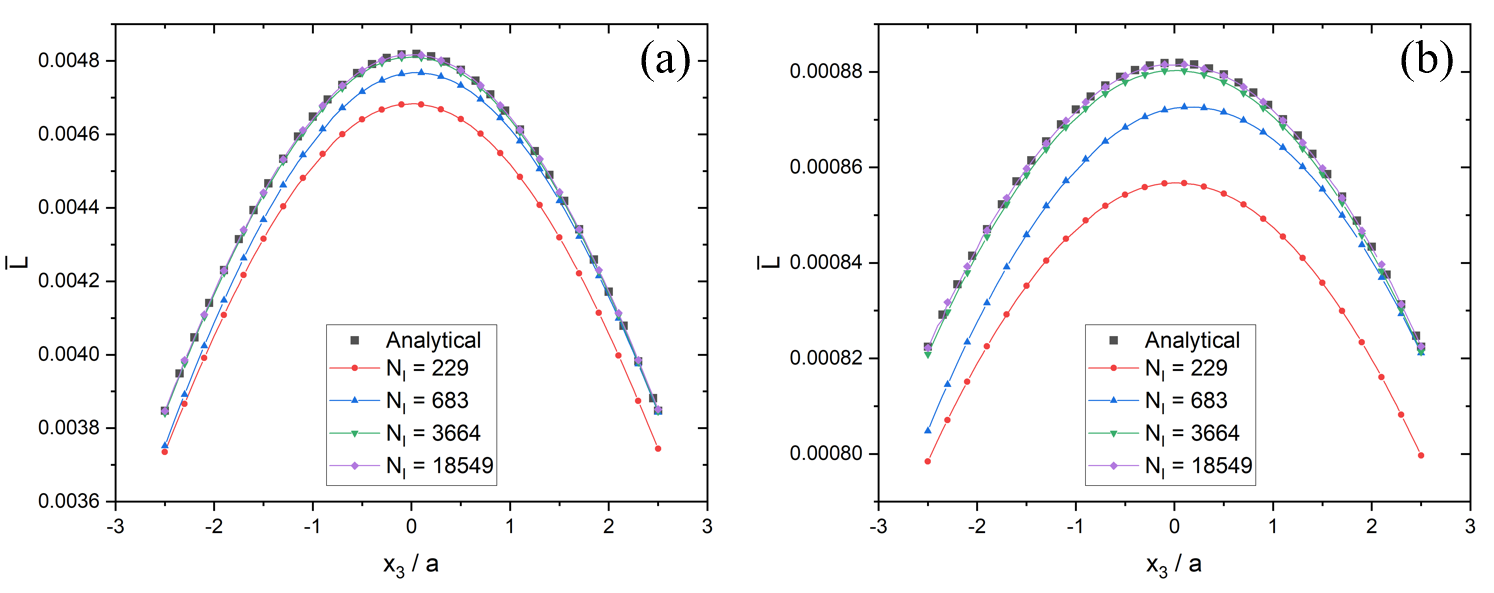}
    \caption{Reproduction and verification of the analytical spherical integral with radius $a = 0.1$ m, $\alpha = 0.05 m^2 / s$ containing a uniform heat source (existing within [0, 1] s), along the vertical center line $x_3 \in [-2.5, 2.5] a$ on (a) different $N_I$-surface polyhedron ($229$, $683$, $3,664$, and $18,549$) with $t = 2 s$; (b) different $N_I$-surface polyhedron with $t = 5 s$. ($n_{max} = 10$)}
    \label{fig:verify_spher_transient_time}
\end{figure}

Subsequently, to verify the time Eshelby's tensor $\overline{L}^{f}(\textbf{x}, t)$, consider a volumetric heat source uniformly distributed within the inclusion $\Omega$, and the heat source exists between time intervals $[0, 1]$ s. The time interval is 1 s for each time station.  Figs. \ref{fig:verify_spher_transient_time} (a) and (b) compare the spatial variation of $\overline{L}$ among analytical solution of Eq. (\ref{eq:3D_time}) when $n = 0$ and solutions approximated by polyhedrons ($N_I = 229$, $683$, $3,664$, and $18,549$). Shown in Fig. \ref{fig:verify_spher_transient_time} (a), when more polyhedrons approximate the sphere, discrepancies between the approximated solution and the analytical solution become smaller. However, when the time intervals increase from 2 s to 5 s, differences become greater among two cases, $N_I = 229$ and $N_I = 683$, with the analytical solution caused by accumulated effects in the spatial integral.

% two figures on Integral and error comparison

\subsubsection{Domain integrals of transient Green's function over the cuboid inclusion}
\begin{figure}
    \centering
    \includegraphics[width = 1\linewidth,height = \textheight,keepaspectratio]{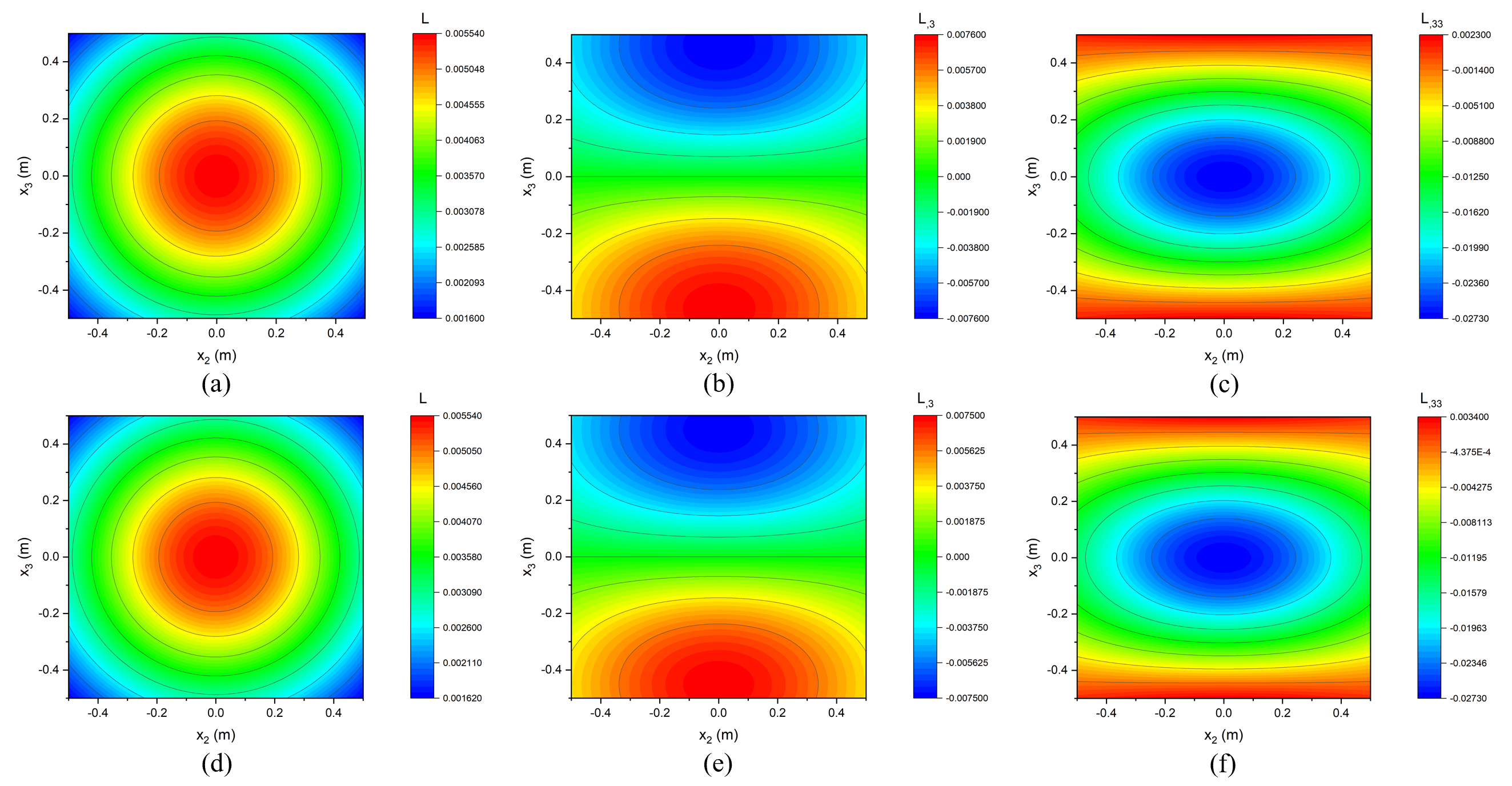}
    \caption{Comparison of domain integrals of transient Green's functions over a cuboid inclusion (a) $L$, (b) $L_{,3}$, (c) $L_{,33}$ by the method of Fourier transform powered with fast-Fourier-transform (FFT), and (d) $L$, (e) $L_{,3}$, (f) $L_{,33}$ by series-form solution. Time $t = 2$ s, $\alpha = 0.05 m^2 / s$, $n = 10$, and $x_2, x_3 \in [-0.5, 0.5]$ m, while $x_1 = 0$. }
    \label{fig:Contour_cuboid}
\end{figure}

Consider a cuboid inclusion with dimension $0.2 \times 0.2 \times 0.2$ m, whose center is located at the origin. Fig. \ref{fig:Contour_cuboid} shows contour plots of domain integrals of transient Green's function over the cuboid inclusion. Without the loss of any generality, since the cuboid is symmetric, this subsection selects $x_2 - x_3$ plane and partial derivatives with respect to $x_3$ for illustration purposes. Figs. \ref{fig:Contour_cuboid} (a-c) are evaluated by the method of Fourier transform, and the characteristic function in the Fourier space is defined as:

\begin{equation}
    \Theta(\textbf{k}) = \int_{-l/2}^{l/2} \int_{-l/2}^{l/2} \int_{-l/2}^{l/2} \exp\left[ -i k_i x_i \right] \thinspace dx_1 dx_2 dx_3 = \frac{8 \sin (\frac{k_1 l}{2}) \sin(\frac{k_2 l}{2}) \sin(\frac{k_3 l}{2})}{k_1 k_2 k_3}
    \label{eq:character_cube}
\end{equation}
where the $l = 0.2$ m is the length. Figs. \ref{fig:Contour_cuboid} (d-f) are calculated by Eqs. (\ref{eq:direct_vol}) and (\ref{eq:stokes_spatial}), and their partial derivative, respectively. Minor discrepancies can be observed among evaluations by two approaches, particularly, Figs. \ref{fig:Contour_cuboid}(a) and (d), which do not involve partial differentiation. Despite the inclusion of cuboid, the distribution of $L$ is center symmetric, and contour lines do not depict the sharp geometry, such as vertices. The distribution pattern is similar to Fig. \ref{fig:verify_spher_transient} (a), which reaches the maximum at the center and decreases with the distance. Fig. \ref{fig:Contour_cuboid} (b) implies that the maximum partial derivative of $L$ does not occur at the center, and the partial derivatives are symmetric to the center line $x_3 = 0$.

An interesting phenomenon can be found in Fig. \ref{fig:Contour_cuboid} that the second order partial derivative $L_{,33}$ is continuous in $x_2$ and $x_3$ directions. Compared to steady-state Eshelby's tensors \cite{Mura1987} and harmonic thermal Eshelby's tensors (see Fig. 4 in \cite{Wu2024-prsa}), these Eshelby's tensors all exhibit discontinuity for interfaces with different normal vectors, i.e., $D_{33}$ is discontinuous at interfaces (with jump $1$, if multiplied with $K$). The interesting phenomenon has been predicted by analytical derivation in Eq. (\ref{eq:L_inverse1}) that the spatial integral of transient Green's function does not require discussing interior/exterior observing points. This phenomenon can be interpreted in two aspects: (a) physical: transient Green's function is governed by the heat equation with a point heat source, i.e., $\delta(\textbf{x} - \textbf{x}') \delta(t-t_0)$. The spatial integrals of transient Green's function will not trigger the interior singularity. The only possibility is a subsequent temporal integral from $t \in [t_0, t]$, where $t$ is any arbitrary time since $t_0$; (b) mathematical: a detailed proof on discontinuities of steady-state/time-harmonic and spatial/time Eshelby's tensors are elaborated in the following subsection.

% Figure compares contour plots: 

\begin{figure}
    \centering
    \includegraphics[width = 1\linewidth,height = \textheight,keepaspectratio]{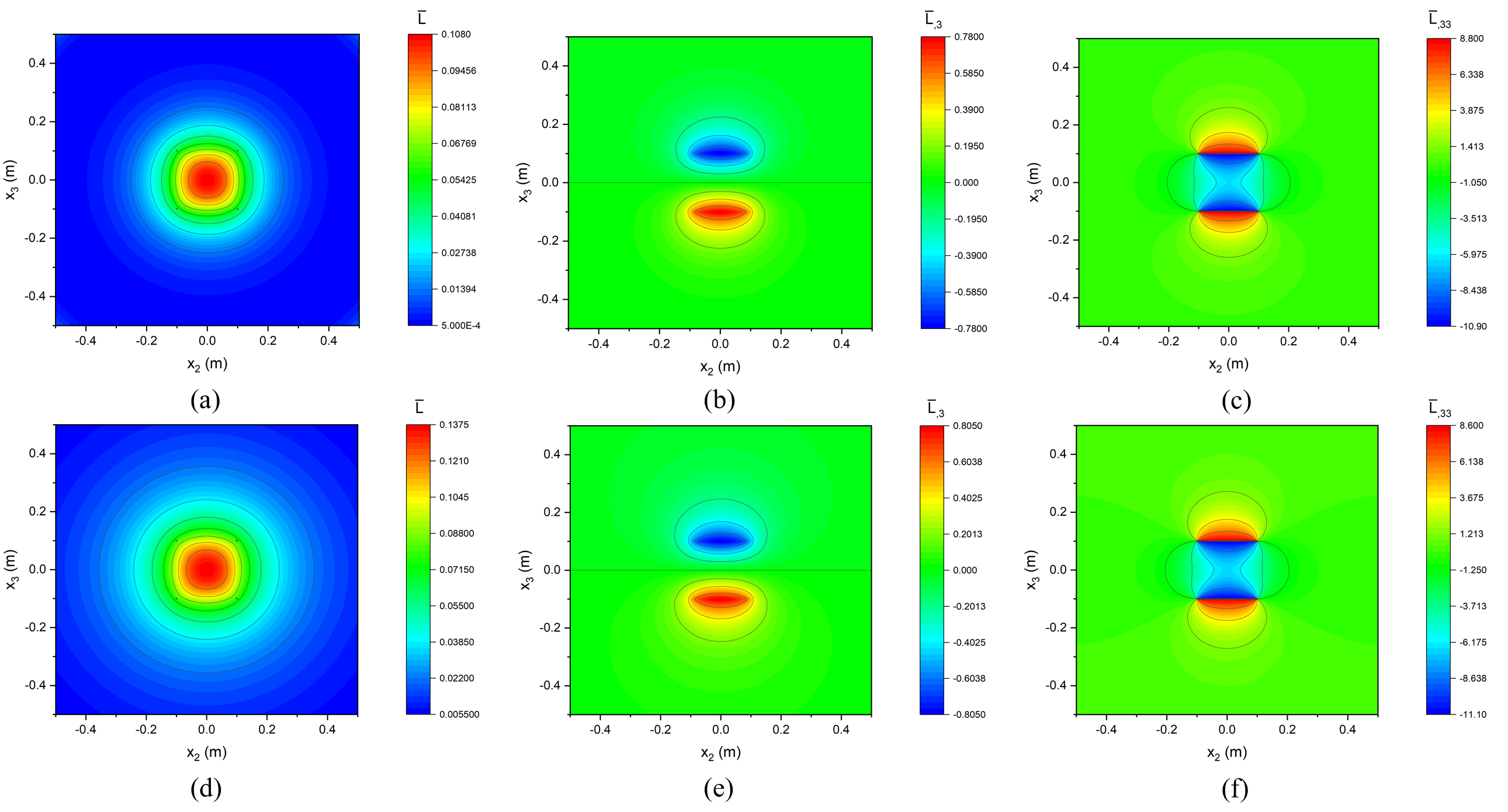}
    \caption{Comparison of spatio-temporal integrals of transient Green's functions over a cuboid inclusion (a) $\overline{L}$, (b) $\overline{L}_{,3}$, (c) $\overline{L}_{,33}$ when time is 0.5 s, and (d) $\overline{L}$, (e) $\overline{L}_{,3}$, (f) $\overline{L}_{,33}$ when time is 5 s. $\alpha = 0.05 m^2 / s$, $n = 10$, and $x_2, x_3 \in [-0.5, 0.5]$ m. }
    \label{fig:Contour_cuboid_time}
\end{figure}

Since two methods of domain integrals produce consistent results, time Eshelby's tensors are plotted in the plane $x_2 - x_3$ at 0.5, 5 s, using Eqs. (\ref{eq:3D_time}) and (\ref{eq:stokes_time}) and their partial derivatives. Comparing Figs. \ref{fig:Contour_cuboid_time} (a) and (d), we can observe: (i) the influential regions of the cuboid enlarges with time, suggesting the heat transfers to further positions provided a longer time; (ii) Eshelby's tensor $\overline{L}$ increases with time, which indicates the heat accumulates in the neighborhood of the inclusion and its distribution becomes more similar to steady-state distribution, see Eqs. (\ref{eq:int_seq_complete}) as the lower incomplete gamma function decreasing with time. Figs. \ref{fig:Contour_cuboid_time} (b) and (e) show that $\overline{L}_{,3}$ is symmetric about its centerline. The similar distribution pattern indicates that the partial derivative of $\frac{1}{\textbf{x} - \textbf{x}'|}$ (steady-state component) dominates. Figs. \ref{fig:Contour_cuboid_time} (c) and (f) support the prior conclusion on discontinuity of Eshelby's tensors $\overline{L}_{,33}$ that Eshelby's tensors are not continuous in $x_2$ and $x_3$ directions. However, it should be noted that we previously concluded that the discontinuity along the direction $x_3$ must be $\frac{1}{K}$, as it represents the physical meaning of the Dirac delta function on point excitations. In Figs. \ref{fig:Contour_cuboid_time} (c) and (f), the differences between maximum (red) and minimum (blue) values (in the color bar) are the same as $19.7 = 0.985 \frac{1}{K}$ because Eshelby's tensors $\overline{L}_{,33}$ do not multiply with $K$. Note that the observing points exhibit some distances to the upper and lower edges of the cube to avoid a numerically unstable jump ($\tan^{-1} [l_f / b]$, if $b \to 0$). However, the $1.5 \%$ difference is observed. Hence, the time Eshelby's tensors will exhibit discontinuities along the interfaces. Particularly when the interface normal vector and partial derivative indices are the same, i.e., $\textbf{n} = (0, 0, \pm1)$, $\overline{L}_{,33}$ and $\overline{D}_{,33}$ exhibit the jump $\frac{1}{K}$ and $1$, respectively. The observations agree well with the derivations in the following subsection. 

% Figure compares contour plots for time Eshelby's tensors: 

\subsection{Discussion of the discontinuity of Eshelby's tensor}
This section explains discontinuities of steady-state, time-harmonic, and spatial/time transient Eshelby's tensors. We first summarize why Eshelby's tensors of steady-state or time-harmonic heat transfer have discontinuity on interfaces with second-order partial derivatives. Without the loss of any generality, second-order partial derivatives of Eshelby's tensors for steady-state and time-harmonic cases can be written in a general-form Green's function $G(\textbf{x}, \textbf{x}')$ as:

\begin{equation}
\begin{split}
    L_{,ij} & = \int_{\Omega} G_{,ij}(\textbf{x}, \textbf{x}') \thinspace d\textbf{x}' = \int_{\Omega} G_{,i'j'}(\textbf{x}, \textbf{x}') \thinspace d\textbf{x}' = -\int_{\partial \Omega} n_i(\textbf{x}') G_{,j} (\textbf{x}, \textbf{x}')\thinspace d\textbf{x}'
\end{split}
\label{eq:Eshleby_two}
\end{equation}
When the source and field points coincide on a smooth boundary, derivatives of Green's function exhibit weak singularities. Following the exclusion technique used in the boundary element method \cite{Beer2008}, Fig. \ref{fig:weak_singularity} shows a small exclusion region with radius $\varepsilon \rightarrow 0$, and source points are moving at the hemispherical shell. In such a case, Eq. (\ref{eq:Eshleby_two}) can be written as superpositions of two integrals: 

\begin{figure}
    \centering
    \includegraphics[width = 0.6\textwidth]{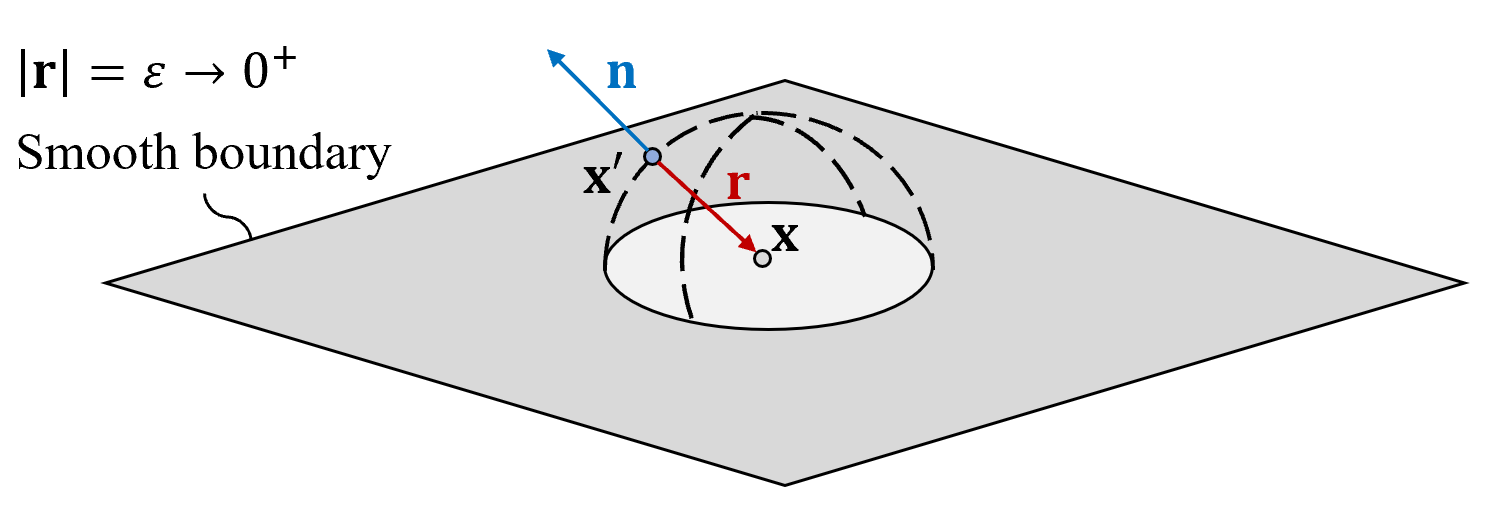}
    \caption{Schematic illustration of an exclusion region with radius $\varepsilon$ ($\varepsilon \to 0$), when the source $\textbf{x}'$ and field point \textbf{x} coincide. $\textbf{r} = \textbf{x} - \textbf{x}'$ and $\textbf{n}$ is the unit outward normal vector of the exclusion boundary}
    \label{fig:weak_singularity}
\end{figure}

\begin{equation}
    \begin{split}
    L_{,ij} & = -\int_{\partial \Omega - \partial \Omega^\varepsilon} n_{i}(\textbf{x}') G_{,j}(\textbf{x}, \textbf{x}') \thinspace d\textbf{x}' - \int_{\partial \Omega^\varepsilon} n_{i}(\textbf{x}') G_{,j}(\textbf{x}, \textbf{x}') \thinspace d\textbf{x}'
\end{split}
\label{eq:Eshleby_two_sep}
\end{equation}
where $\partial \Omega$ and $\partial \Omega^{\varepsilon}$ refers to the boundary and a small exclusion of the boundary, respectively; the first integral does not change when the field point moves from the inner to the outer surface of the boundary; however, the second integral is the opposite for inner and outer field points because the dot product of distance vector $\textbf{r}$ and normal vector $\textbf{n}$ changes the sign. Therefore, the discontinuity can be obtained by doubling the second integral for the field point, as the normal vector and the distance vector are opposite, $n_i r_i / (|\textbf{n}| |\textbf{r}|) = -\sin \psi$). Substituting the first partial derivative of Green's function into Eq. (\ref{eq:Eshleby_two_sep}) with $\textbf{x}^+$ and $\textbf{x}^-$ representing field points on the outer and inner surface, we can obtain:  

\noindent (1) For steady-state heat conduction
\begin{equation}
    L_{,ij}(\textbf{x}^+) - L_{,ij}(\textbf{x}^-) = 2 \int_{0}^{2\pi} \int_{0}^{\frac{\pi}{2}} \frac{1}{4 \pi K} \frac{1}{\varepsilon^2} \sin\psi \varepsilon^2 d\psi d \thinspace \theta = \frac{1}{K}
\end{equation}

\noindent (2) For time-harmonic heat transfer
\begin{equation}
    L_{,ij}(\textbf{x}^+) - L_{,ij}(\textbf{x}^-) = 2 \int_{0}^{2\pi} \int_{0}^{\frac{\pi}{2}} \left( \frac{\exp[i \beta \varepsilon]}{4 \pi K \varepsilon^2} - \varepsilon \beta \frac{\exp[i \beta \varepsilon]}{4 \pi K \varepsilon^2} \right) \sin\psi \varepsilon^2 d\psi d \thinspace \theta = \frac{1}{K}
\end{equation}

\noindent (3) For spatial Eshelby's tensor for transient heat transfer at $t>0$:
\begin{equation}
    L_{,ij}(\textbf{x}^+, t) - L_{,ij}(\textbf{x}^-, t) = \frac{2}{C_p} \int_{0}^{2\pi} \int_{0}^{\frac{\pi}{2}} \varepsilon \frac{\exp[\frac{-\varepsilon^2}{4 \alpha t}]}{16 \pi^{\frac{3}{2}} (\alpha t)^\frac{5}{2}} \sin\psi \varepsilon^2 d\psi d \thinspace \theta = 0
\end{equation}

\noindent (4) For time Eshelby's tensor for transient heat transfer, we can interchange the sequence of spatial and temporal integral. When $t_i > 0$, the integral yields:

\begin{equation}
\begin{split}
    L_{,ij}(\textbf{x}^+, t) - L_{,ij}(\textbf{x}^-, t) & = \frac{2}{C_p}\int_{0}^{2\pi} \int_0^{\frac{\pi}{2}} \frac{1}{4 \pi^{\frac{3}{2}} \alpha^{\frac{3}{2}} (t-t_f)^\frac{1}{2} \varepsilon^2} \left\{ \exp \left[\frac{-\varepsilon^2}{\alpha (t - t_f)}\right] - \exp \left[\frac{-\varepsilon^2}{\alpha (t - t_{f-1})}\right] \right\} \varepsilon^2 \thinspace \sin \psi \thinspace d\psi \thinspace d\theta \\ 
    & + \frac{2}{C_p} \int_{0}^{2\pi} \int_0^{\frac{\pi}{2}} \frac{1}{4 \pi^{\frac{3}{2}} \alpha \varepsilon^2} \left\{ \Gamma\left[ \frac{1}{2}, \frac{\varepsilon^2}{4 \alpha (t - t_f)} \right] - \Gamma\left[ \frac{1}{2}, \frac{\varepsilon^2}{4 \alpha (t - t_{f-1})} \right] \right\} \varepsilon^2 \thinspace \sin \psi \thinspace d\psi \thinspace d\theta = 0
\end{split}
\end{equation}

When $t_ i = 0$, the incomplete Gamma function should be altered as Eq. (A.2),

\begin{equation}
\begin{split}
    L_{,ij}(\textbf{x}^+, t) - L_{,ij}(\textbf{x}^-, t) & = \frac{2}{C_p} \int_{0}^{2\pi} \int_{0}^{\frac{\pi}{2}} \frac{1}{4 \pi \alpha} \frac{1}{\varepsilon^2} \sin\psi \varepsilon^2 d\psi d \thinspace \theta + \frac{2}{C_p} \int_{0}^{2\pi} \int_{0}^{\frac{\pi}{2}} \frac{1}{4 \pi^{\frac{3}{2}} \alpha^2 } \left\{ \frac{\exp\left[ \frac{-\varepsilon^4}{16 \alpha^2 (t-t_{f-1})^2} \right]}{t - t_{f-1}} \right. \\ & \left. - \frac{\sqrt{\pi} \alpha \text{Erf} \left[ \frac{\varepsilon^2}{4 \alpha (t-t_{f-1})} \right]}{\varepsilon^2} \right\} \varepsilon^2 \sin \psi \thinspace d \psi \thinspace d\theta = \frac{1}{K}
\end{split}
\end{equation}
The above derivations are consistent with previous results in the literature for steady-state Eshelby's tensor \cite{Hatta1986}, time-harmonic Eshelby's tensor \cite{Wu2024-prsa}, and spatial/time Eshelby's tensor reported in the current paper. 

\subsection{Extension to the spheroidal inhomogeneity problem}
As the domain integrals in the series-form and Fourier space have been verified by the closed-form spherical time/spatial Eshelby's tensors, a more practical application of inclusion problems is the equivalent inclusion method (EIM), which replaces the original inhomogeneity by the matrix with continuously distributed eigen-fields. Since polygonal/polyhedral inclusions have angular vertices, its disturbed local fields (heat flux) are singular at the vertices, even for a uniformly distributed ETG. Wu and Yin \cite{Wu2024_JMPS} has proved that the eigen-field for an angular inhomogeneity is singular, which is a primary limitation preventing from implementation of the polynomial-form EIM on angular inhomogeneities. 

Although there is a limitation of the current polynomial-form EIM, extensions can be made using cuboid inclusions assuming uniform eigen-fields. The pioneering work in semi-infinite elastostatics \cite{Liu2012} provides insights on the utilization of numerical Fourier-transform to solve inhomogeneities decomposed into many elementary cubes, which can be the future development as this paper mainly aims to convey derivation of Eshelby's tensors instead. Therefore, this section only focuses on the spheroidal inhomogeneity problem (without closed-form Eshelby's tensors) for demonstration of the method. Note that the (multiple) spherical inhomogeneity problem (with closed-form Eshelby's tensors) was solved in \cite{Wu_IJSS_2025}, which will not be repeated below.

\subsubsection{Equivalent conditions}
When the inhomogeneity exhibits different thermal conductivity and thermal capacity from the matrix, Wu and Yin \cite{Wu_IJSS_2025} proposed to treat it as an equivalent inclusion with material constants ($K^0, C_p^0$) containing continuously distributed eigen-fields (EHS and ETG). The EHS and ETG are determined by the equivalent flux and heat source conditions, which ensure the equivalent inclusion exhibits the same heat flux and energy storage/release rate as the original inhomogeneity problem: 
%% polynomial-form equivalent conditions
\begin{equation}
\begin{aligned}
    K^0 (u_{i}^0 + u_{i}' - u^{0*}_{i}) & = K^I (u_{i}^0 + u_{i}') \\ 
    K^0 (u_{i,p}^0 + u_{i,p}' - u^{1*}_{ip}) & = K^I (u_{i,p}^0 + u_{i,p}') \\
    K^0 (u_{i,pq}^0 + u_{i,pq}' - 2 u^{2*}_{ipq}) & = K^I (u_{i,pq}^0 + u_{i,pq}')
\end{aligned}
    \label{eq:equiv_ETG}
\end{equation}
which is applicable to all heat conduction problems at the center of the inhomogeneity. In Eq. (\ref{eq:equiv_ETG}), the superscript $0$ represents undisturbed fields (without the inhomogeneity). For general transient conduction problems, the equivalent heat source conditions are additionally required as,
\begin{equation}
    \begin{aligned}
    C_p^0 \frac{\partial u}{\partial t} & = C_p^I \frac{\partial u}{\partial t} - Q^{0*} \\
    C_p^0 \frac{\partial^2 u}{\partial t \partial x_p} & = C_p^I \frac{\partial^2 u}{\partial t \partial x_p} -Q^{1*}_{p} \\
    C_p^0 \frac{\partial^3 u}{\partial t \partial x_p \partial x_q} & = C_p^I \frac{\partial^3 u}{\partial t \partial x_p \partial x_q} - 2 Q^{2*}_{pq} \\
    \end{aligned}
    \label{eq:equiv_EHS}
\end{equation}
The combination of Eqs. (\ref{eq:equiv_ETG}) and  (\ref{eq:equiv_EHS}) constructs a system of linear equations, whose solutions are ETG and EHS. Based on previous assumption of uniform, linear and quadratic approximations, the polynomial-form EIM requires $4, 16$ and $52$ unknowns to be determined, respectively. And the details of matrix arrangements to determine eigen-fields have been explained in the global matrix Eq. (29) in \cite{Wu_IJSS_2025}. 

\subsubsection{Handling on boundary/initial conditions}
Unlike the verification of steady-state problems, the simulation of transient heat conduction requires proper setting of boundary/initial conditions. In general, the accuracy of undisturbed temperature gradient $u_{i}^0$ significantly affect EIM's results. Following a classic boundary integral equations by \cite{Brebbia1984, Gupta1995}, the temperature at any interior point with uniform initial prescribed temperature, 

\begin{fontsize}{8}{10}
\begin{equation}
u^0(\textbf{x}, t)  =  \int_{\partial \mathcal{D}} \int _{\Lambda} \alpha(\textbf{x}') G(\textbf{x}, \textbf{x}', t, t') q^{BC}(\textbf{x}') \  dt' \thinspace d\textbf{x}' - \int_{\partial \mathcal{D}} \int _{\Lambda} \alpha(\textbf{x}') T(\textbf{x}, \textbf{x}', t, t') u^{BC}(\textbf{x}') \ dt' \thinspace d\textbf{x}' + \int_{\mathcal{D}} \int_{\Lambda} G(\textbf{x}, \textbf{x}', t, t') u^{I}(\textbf{x}') \thinspace dt' \thinspace d\textbf{x}'
\label{eq:BIE}
\end{equation}
\end{fontsize}
where $u^{BC}$ and $q^{BC}$ represents the temperature and flux on the boundary; $T(\textbf{x}, \textbf{x}', t, t') = -K \frac{\partial G(\textbf{x}, \textbf{x}', t, t')}{\partial x_i'} n_i (\textbf{x}')$ is the fundamental solution of flux; $\partial \mathcal{D}$ and $\Lambda$ denote surface and time integral domain, respectively; and $u^{I}(\textbf{x})$ is the initial prescribed temperature. Without the loss of any generality, $u^I(\textbf{x}) = 0^\circ C$ is applied, and thus the last integral term vanishes. Since a transient fundamental solution shares similar features as the static case in which the strongly singular kernel function $T$ decays as $1/r^2$, where $r$ is the distance. Our recent book \cite{Yin_iBEM} has demonstrated that when one inhomogeneity is $5$ times away from a field point, its influence on that field point is small enough can be neglected with tolerance. The temperature field is the summation of Eqs. (\ref{eq:BIE}) and (\ref{eq:super_disturb_time}), which considers contribution from boundary/initial conditions and inhomogeneity's eigen-fields, respectively.

\subsubsection{Comparison with FEM results}
\begin{figure}
    \centering
    \includegraphics[width = 0.3 \textwidth,keepaspectratio]{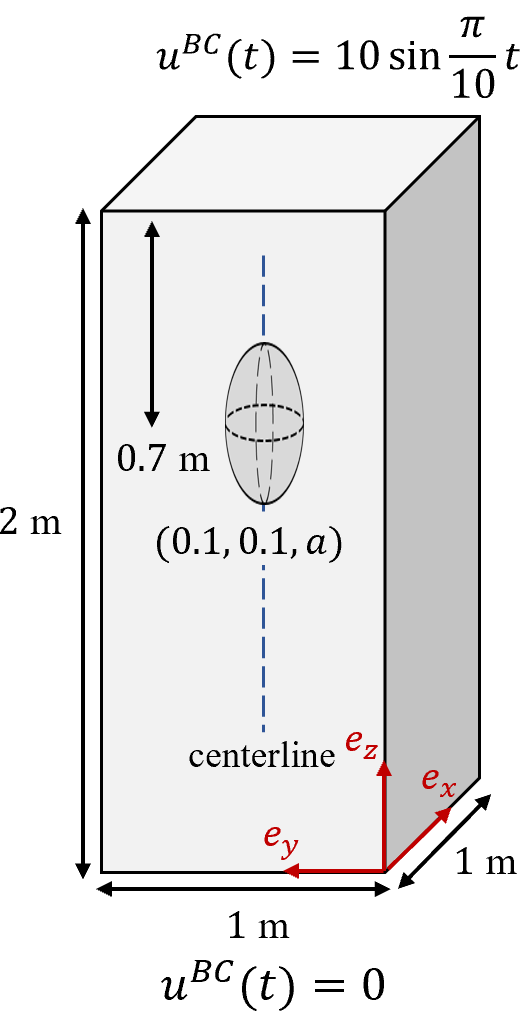}
    \caption{Schematic plot of one block containing a spheroidal inhomogeneity with dimension 1 $\times 1 \times 2$ m, which is subjected to prescribed sinusoidal and constant temperature load on its top and down surfaces, respectively}
    \label{fig:5}
\end{figure}

Fig. \ref{fig:5} shows the set up of the inhomogeneity problem that a single inhomogeneity is located at $(0.5, 0.5, 1.3)$ m. The boundary conditions are (a) sinusoidal temperature variation $10 \sin \frac{\pi}{10} t$ is applied on the top surface; (b) constant zero temperature on the bottom surface; and (c) all side surfaces are insulated. Three semi-axes of the spheroidal inhomogeneity are $(0.1, 0.1, a)$ m, where $a = 0.06, 0.12$ m, respectively. And the transient thermal properties are $K^0 = 1 W / m K, K^1 = 10 W / m K$ and $C_p^0 = 10 J / m^3 K, C_p^1 = 12 J / m^3 K$.

The sinusoidal load has a period of 20 s, and the verification case compares results within the first half period (10 s). To ensure a fair comparison in computational time between FEM and EIM, both the FEM (ANSYS) and EIM utilizes the same time interval of 0.05 s (200 time steps). It should be noted the current time discretization leads to some fluctuations around the inhomogeneity for FEM, while the EIM maintain smoother solutions. (a) For FEM, when $a = 0.06, 0.12$ m, it employs $229, 240$, $201, 738$ nodes and $164,795$, $144, 429$ 10-node tetrahedral elements, respectively. And two cases takes $491$, $439$ s, respectively. (b) For EIM, the boundary integral equation requires $1,000$ 4-node quadrilateral boundary elements and $1,002$ boundary nodes, and the uniform, linear, and quadratic takes $109, 116, 145$s, respectively. Keep in mind that the time consumption of EIM is longer than the spherical case study (73 s), because it requires numerical inverse transform procedures, as the spheroidal inclusion does not have explicit time Eshelby's tensors in the temporal-spatial space. 

% Figures of a = 0.06m 
\begin{figure}
    \centering
    \includegraphics[width = 1 \linewidth, keepaspectratio]{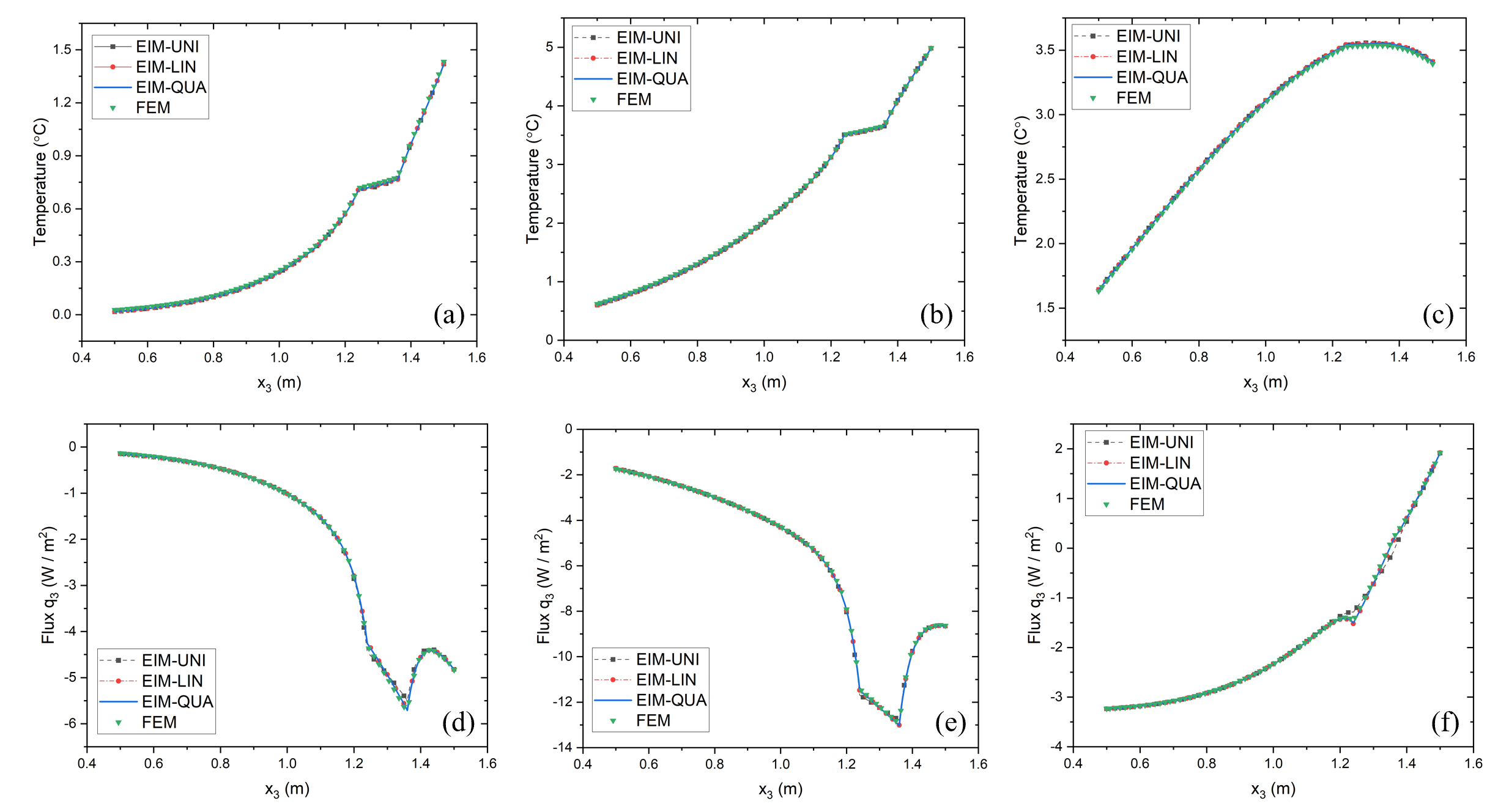}
    \caption{Variation and comparison of thermal fields caused by a spheroidal inhomogeneity (0.1, 0.1, 0.06) m through EIM with uniform, linear and quadratic eigen-fields and FEM along the the vertical center line ($x_3 \in [0.5, 1.5]$ m) at different time, (a) temperature at 2 s; (b) temperature at 5 s; (c) temperature at 10 s; (d) heat flux at 2 s; (e) heat flux at 5 s; and (f) heat flux at 10 s.}
    \label{fig:a06}
\end{figure}

% Figures of a = 0.12m
\begin{figure}
    \centering
    \includegraphics[width = 1 \linewidth, keepaspectratio]{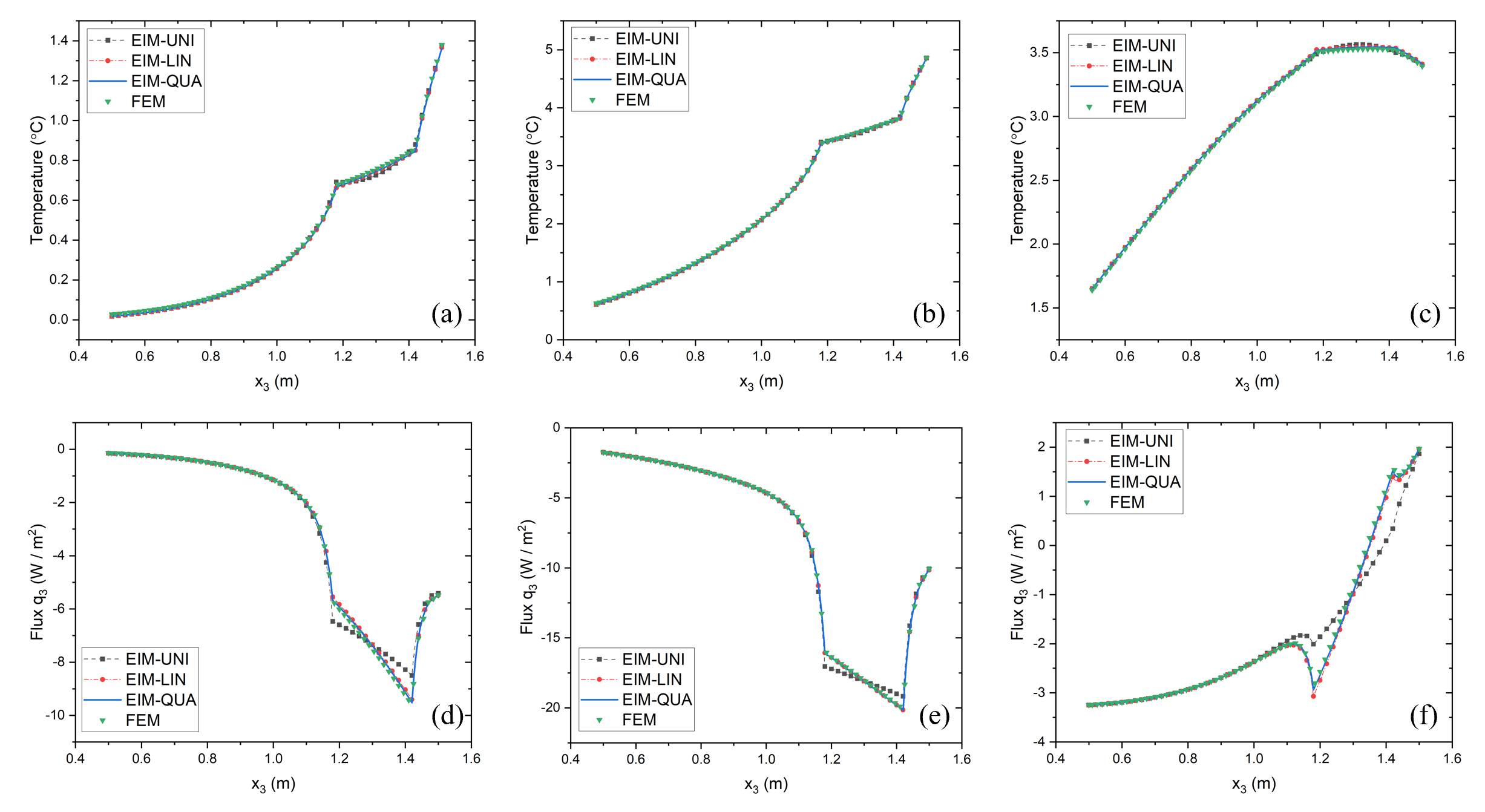}
    \caption{Variation and comparison of thermal fields caused by a spheroidal inhomogeneity (0.1, 0.1, 0.12) m through EIM with uniform, linear and quadratic eigen-fields and FEM along the the vertical center line ($x_3 \in [0.5, 1.5]$ m) at different time, (a) temperature at 2 s; (b) temperature at 5 s; (c) temperature at 10 s; (d) heat flux at 2 s; (e) heat flux at 5 s; and (f) heat flux at 10 s.}
    \label{fig:a12}
\end{figure}

Fig. \ref{fig:a06} plots the verification and comparison of temperature and heat flux along the vertical center line ($x_3 \in [0.5, 1.5]$ m), when the third semi-axis of the spheroid is 0.06 m. Because the top surface exhibits higher temperature, the undisturbed temperature gradually decreases from the top to the bottom, which leads to nonzero temperature gradients. As indicated in Eqs. (\ref{eq:equiv_ETG}) and (\ref{eq:equiv_EHS}), the nonzero temperature gradient causes ETG and linearly distributed EHS. In such a case, the uniform assumption on EHS can seldom provide accurate predictions. 

Figs. \ref{fig:a06} (a) and (b) compare temperature fields at $t = 2$ and $t = 5$ s, respectively. Although the uniform assumption of eigen-fields is not accurate, the curve ''EIM-UNI'' can provide acceptable predictions for observing points with distance more than $1.2 a$. Its accuracy rapidly decreases when observing points are close or interior the inclusion, because spatial-variation of eigen-fields play a more significant role in the close neighborhood of the equivalent inclusion. However, curves with linear and quadratic assumption of eigen-fields exhibit small discrepancies compared with temperature by the FEM, especially for interior observing points. Such phenomenon suggests that when $t$ is small, uniform and linear terms eigen-fields dominates, and so does the temperature gradient. Observing the prescribed temperature at the top surface, the sine function generally exhibits greater derivatives at smaller time steps. Figs. \ref{fig:a06} (c-f) plot the corresponding heat flux along the center line when $t = 2$, $t = 5$ and $t = 10$ s. The curve ``EIM-UNI'' exhibits minor discrepancies between other curves, even in the neighborhood of interfaces of inhomogeneities. One reason is that the although the equivalent conditions are only conducted at the center of the inhomogeneity, the third axis of the spheroid is only 0.06 m, which is comparatively smaller than the two other axes. Another reason is that, when the time increases (within the quarter period), the undisturbed temperature gradient decreases, which results in a smaller ETG and EHS. Such decrease subsequently leads to smaller linear terms of the approximating polynomials. Fig. \ref{fig:a06} (d) and Fig. \ref{fig:a06} (e) support our previous conclusions of effects of undisturbed temperature gradient on eigen-fields. Subsequently, when time $t$ become $10$ s in Figs. \ref{fig:a06} (f), increasing discrepancies can be found compared to Fig. \ref{fig:a06} (d). Therefore, even the eigen-fields may vary along the gradient (z-axis) direction, the influences of the inaccurate eigen-fields do not play a significant role. Similar to temperature fields, such influences rapidly decreases with distance for exterior observing points. The comparison of temperature and heat flux at different time (change of temperature gradient) clearly demonstrates that the spatial variation of undisturbed temperature gradients can significantly influence predictions by the EIM with uniform assumption of eigen-fields, but the inaccurate errors vanishes rapidly for far exterior observing points.

However, the good performances by EIM with uniform assumption of eigen-fields for a spheroid with shorter third axis (a = 0.06 m) does not guarantee accuracy for the case with longer third axis (a = 0.12 m) in Figs. \ref{fig:a12} (a-f). Keep in mind that although the size of inhomogeneity changes, the undisturbed thermal fields retain. Based on our previous analysis, when time is small, i.e., t = 2 s, the temperature gradients are comparatively greater, which lead to eigen-fields with larger magnitude. Moreover, because the spheroid inhomogeneity exhibits a longer third axis, the zeroth order (uniform) equivalent conditions conducted at the center is not able to capture spatial variations of eigen-fields accurately. Therefore, much more apparent discrepancies among the ``EIM-UNI'' and other curves can be found in Figs. \ref{fig:a12} (a, c, d, f), if compared to Figs. \ref{fig:a06} (a,c,d,f). Certainly, the discrepancies in temperature fields are smaller than these of heat flux, which should be attributed two primary reasons. Firstly, heat flux is a higher-order field, and thus its numerical accuracy is more sensitive to eigen-fields. In contrast, the temperature field is obtained with less partial derivative. Secondly, thermal fields are calculated by time convolution, and numerical errors are accumulated by inaccurate solutions with the uniform eigen-fields. Therefore, it is recommended to consider spatial variation of eigen-fields for transient inhomogeneity problems. 

\section{Conclusions}
The local disturbance by continuously distributed source terms, eigen-temperature-gradient and eigen-heat-source, are formulated as domain integrals of Green's functions over the inclusion, which are defined as generalized Eshelby's tensors. This paper derives generalized Eshelby's tensors for the polynomial form of source fields on 2D polygonal and 3D polyhedral inclusions, and applied the Fourier transform to derive Eshelby's tensors over ellipsoidal inclusions. Closed-form Eshelby's tensors are provided for spherical inclusion with polynomial-form eigen-fields. The discontinuity of spatial and time Eshelby's tensors are presented and discussed, which was later supported by case study of a cuboid inclusion. The proposed series-form Eshelby's tensors are verified against Michelitsch's analytical solution for the uniform eigen-field in the harmonic state, as well as our analytical solution for the uniform, linear, quadratic eigen-fields in the transient state. Further verifications were conducted through comparisons with numerical solutions evaluated through the fast Fourier transform. The generalized Eshelby's tensors provide a foundational framework for future studies in this direction. While these findings advance the understanding for local disturbances in both harmonic and transient states, the authors acknowledge that further extensions and developments may be required to address more complex inhomogeneity problems.

\section*{‌CRediT Author Contribution}
\textbf{Chunlin Wu}: Conceptualization, Methodology, Data Curation, Software, Validation, Writing-Original Draft, Funding Acquisition; \textbf{Zhenhua Wei}: Writing-Review \& Editing; \textbf{Huiming Yin}: Writing-Review \& Editing, Supervision.

\appendix 
\section{Derivation of Eshelby's tensors for harmonic heat transfer} \label{appendix:harmonic}

When the heat source follows a harmonic function with an  excitation frequency $\omega$, as $u(\textbf{x}, t) = \overline{u}(\textbf{x}) \exp[-i \omega t]$, the heat equation can be changed to a modified Helmholtz's equation \cite{Wu2024-prsa} as: 

% governing equation in time-harmonic domain
\begin{equation}
     -\nabla^2 \overline{u}(\textbf{x}) - \frac{i \omega}{\alpha} \overline{u} (\textbf{x}) = \frac{\overline{Q}(\textbf{x})}{K}
     \label{eq:harmonic}
\end{equation}
The fundamental solution of Eq. (\ref{eq:harmonic}) can be obtained by the Fourier transform on Eq. (\ref{eq:fund_time}) from time domain to the frequency domain as:
\begin{equation}
    G^H(\textbf{x}, \textbf{x}',\beta) = \frac{1}{C_p} \int_{-\infty}^{\infty} \exp[i \omega t] (4 \pi \alpha t)^{\frac{-n}{2}} \exp[\frac{-|\textbf{x} - \textbf{x}'|^2}{4 \alpha t}] H(t) \thinspace d t = \frac{1}{4 K} \begin{cases} 
    i H_0^{(1)} \left[ \beta |\textbf{x} - \textbf{x}'| \right] & \text{n = 2} \\
    \frac{\exp[i \beta |\textbf{x} - \textbf{x}'|]}{\pi|\textbf{x} - \textbf{x}'|} & \text{n = 3}
    \end{cases}
    \label{eq:fund_harmonic}
\end{equation}
where $\beta = \sqrt{\frac{i\omega} {\alpha}}=\frac{1+i} {\sqrt 2}\sqrt{\frac{\omega} {\alpha}}$ is a complex parameter, which is similar to the definition of the wave number in acoustics; and $H_0^{(1)}$ is the Hankel function of the first kind. When $\omega$ or $\beta$ approaches zero, the above can be reduced to the steady-state fundamental solution. The same domain integrals can be applied over inclusions as Helmholtz's potential. For harmonic eigen-fields, polynomial-form Eshelby's tensors are provided as follows \cite{Wu2024-prsa}: 
\begin{equation}
   L^H_{pq...}(\textbf{x})  = \frac{1}{4 \pi K} \Phi_{pq...} (\textbf{x}) \quad \text{and} \quad  D^H_{ipq...}(\textbf{x})  = \frac{-1}{4 \pi}  \Phi_{pq...,i} (\textbf{x})
\label{eq:harmonic_eshe}
\end{equation}
where
\begin{equation}
    \Phi_{pq...}(\textbf{x}) = \int_{\Omega} \frac{\exp[i \beta |\textbf{x} - \textbf{x}'|]}{|\textbf{x} - \textbf{x}'|} x'_p x'_q ...  \thinspace d\textbf{x}'
\label{eq:polynomial_Phi}
\end{equation}

Since harmonic solutions can be obtained through a forward Fourier transform of the solution in the time domain, therefore, polynomial-form harmonic Eshelby's tensor can be obtained through the Fourier transform. Note that the details of the Fourier transform have been provided in our recent work \cite{Wu2024-prsa}, only results are listed as follows,
\begin{equation}
\begin{aligned}
    \Phi(\textbf{x}) & = \mathcal{A}^0(\textbf{x})\\ 
     \Phi_{p}(\textbf{x}) &= \frac{i}{\beta} \mathcal{A}^1_{,p}(\textbf{x}) + x^C_p \mathcal{A}^0( \textbf{x})  \\
     \Phi_{pq}(\textbf{x}) &= -\frac{1}{\beta^2} \mathcal{A}^2_{,pq}(\textbf{x}) - \frac{i}{\beta^3} \mathcal{A}^1_{,pq}(\textbf{x}) + \frac{i}{\beta} \big( \delta_{pq} \mathcal{A}^1(\textbf{x}) + x_p \mathcal{A}^1_{,q}(\textbf{x}) + x_q \mathcal{A}^1_{,p}(\textbf{x}) \big) + x_p x_q \mathcal{A}^0(\textbf{x})
\end{aligned}
    \label{eq:Helmholtz_integ}
\end{equation}
where
\begin{equation}
    \mathcal{A}^n(\textbf{x}) = \int_{\Omega} \frac{\exp \left[ i \beta |\textbf{x} - \textbf{x}'| \right]}{|\textbf{x} - \textbf{x}'|} |\textbf{x} - \textbf{x}'|^n \thinspace d \textbf{x}'  
    \label{eq:mathcal_A_n}
\end{equation}
Substituting $\Phi(\textbf{x})$, $\Phi_{p}(\textbf{x})$, and $\Phi_{pq}(\textbf{x})$ into Eq. (\ref{eq:harmonic_eshe}) yields the polynomial-form harmonic Eshelby's tensors. 

\subsection{Derivation of the harmonic Eshelby's tensor}
Note that Wang et al. \cite{Wang2005} first derives Helmholtz's potential over an arbitrarily-shaped domain, and the authors utilize the convolution property of the Fourier space. Their results are only presented as a triple numerical integral (inverse Fourier transform). Alternatively, this subsection provides a series-form domain integral, which does not contain any integral-form expression. 

Following Eq. (\ref{eq:Taylor_fund}), the volume integrals of $\frac{\exp \left[ i \beta |\textbf{x} - \textbf{x}'| \right]}{|\textbf{x} - \textbf{x}'|} |\textbf{x} - \textbf{x}'|^n$ is written as $\mathcal{A}^{n} %Please double check if $\mathcal{A}^{0} with  _,i; No, it is a volume integral. 
(\textbf{x}) = \int_{\Omega} \frac{\exp \left[ i \beta |\textbf{x} - \textbf{x}'| \right]}{|\textbf{x} - \textbf{x}'|} |\textbf{x} - \textbf{x}'|^n \thinspace d\textbf{x}'$, which can be calculated as follows: 

\begin{equation}
\begin{split}
   \mathcal{A}^n(\textbf{x}) & = \sum_{m=0}^{\infty} \frac{(i \beta)^m}{m! (n + m + 1)} \sum_{I=1}^{N_I} \sum_{J=1}^{N_{JI}} \Big\{ -a_I^{m+n} |a_I| \left( \tan^{-1}  \left[ \frac{l_{JI}^+}{b_{JI}} \right] - \tan^{-1}  \left[ \frac{l_{JI}^-}{b_{JI}} \right]  \right) + \frac{(a_I^2 + b_{JI}^2)^{\frac{n + m + 1}{2}}}{b_{JI}}  \\ & \Big( l_{JI}^+ F_1 \left[ \frac{1}{2}, -\frac{n + m + 1}{2}, 1, \frac{3}{2}, \frac{-(l_{JI}^+)^2}{a_I^2 + b_{JI}^2}, \frac{-(l_{JI}^+)^2}{b_{JI}^2} \right] - l_{JI}^- F_1 \left[ \frac{1}{2}, -\frac{n + m + 1}{2}, 1, \frac{3}{2}, \frac{-(l_{JI}^-)^2}{a_I^2 + b_{JI}^2}, \frac{-(l_{JI}^-)^2}{b_{JI}^2} \right] \Big) \Big\}
\end{split}
\label{eq:Helmholtz_direct}
\end{equation}

% \subsubsection{Gauss's and Stoke's theorem on the volume integral} 
Replacing the target function $G(a_I, b_{JI}, le, t, t')$ as $\frac{\exp \left[ i \beta \sqrt{a_I^2 + b_{JI}^2 + le^2} \right]}{\sqrt{a_I^2 + b_{JI}^2 + le^2}} (a_I^2 + b_{JI}^2 + le^2)^{n / 2}$ in Eq. (\ref{eq:Gao_19_cpl}), one can obtain the corresponding functions $\mathcal{G}^{n}(a_I, b_{JI}, le)$ ($n = 0, 1, 2$) as, 
\begin{equation}
    \begin{aligned}
    & \mathcal{G}^{0}(a_I, b_{JI}, le) = \frac{-i}{\beta \sqrt{b^2 + le^2} } \Big( \exp \left[ i \beta \sqrt{a_I^2 + b_{JI}^2 + le^2} \right] - \exp \left[ i \beta |a_I| \right] \Big) \\
    & \mathcal{G}^{1}(a_I, b_{JI}, le) = \frac{1}{\beta^2 \sqrt{b^2 + le^2} } \Big( (1 - i \beta \sqrt{a_I^2 + b_{JI}^2 + le^2}) \exp \left[ i \beta \sqrt{a_I^2 + b_{JI}^2 + le^2} \right]  - (1 - i \beta |a_I|) \exp \left[ i \beta |a_I| \right] \Big) \\
    & \mathcal{G}^{2}(a_I, b_{JI}, le) = \frac{i}{\beta^3 \sqrt{b^2 + le^2} } (2 - 2 i \beta \sqrt{a_I^2 + b_{JI}^2 + le^2} - \beta^2 (a_I^2 + b_{JI}^2 + le^2)) \exp \left[ i \beta \sqrt{a_I^2 + b_{JI}^2 + le^2} \right] 
    \\ & \quad \quad \quad \quad \quad \quad \quad - (2 - 2 i \beta |a_I| - \beta^2 a_I^2) \exp \left[ i \beta |a_I| \right]
    \end{aligned}
\end{equation}

The first-order partial derivatives of $\mathcal{A}^{n}$ ($n = 0, 1, 2$) can also be derived as follows: 
\begin{equation}  
\begin{aligned}
    & \mathcal{A}^{0}_{,i}(\textbf{x}) = \frac{i}{\beta} \sum_{I=1}^{N_I}  (\xi_I^0)_i \sum_{J=1}^{N_{IJ}} \Bigg\{ \sum_{m=0}^\infty \frac{(i \beta)^m}{m!} \left[ C^{m}_{0}(a_I, b_{JI}, l_{JI}^+) - C^{m}_{0}(a_I, b_{JI}, l_{JI}^-) \right] \\ & \quad \quad \quad \quad  - \exp \left[ i \beta |a_I| \right] \left( \tan^{-1} \left[ \frac{l_{JI}^+}{b_{JI}} \right] - \tan^{-1} \left[ \frac{l_{JI}^-}{b_{JI}} \right] \right) \Bigg\} \\
     &\mathcal{A}^{1}_{,i}(\textbf{x}) = \frac{-i}{\beta} \mathcal{A}^0_{,i}(\textbf{x}) + \frac{i}{\beta} \sum_{I=1}^{N_I} (\xi_I^0)_{i} \sum_{J = 1}^{N_{JI}} \Bigg\{ \sum_{m=1}^{\infty} \frac{(i \beta)^m}{m!} \left[ C^{m}_{1}(a_I, b_{JI}, l_{JI}^+) - C^{m}_{1}(a_I, b_{JI}, l_{JI}^-) \right] \\ & \quad \quad \qquad - |a_I| \exp \left[ i \beta |a_I| \right] \left( \tan^{-1} \left[ \frac{l_{JI}^+}{b_{JI}} \right] - \tan^{-1} \left[ \frac{l_{JI}^-}{b_{JI}} \right] \right) \Bigg\} \\
     & \mathcal{A}^{2}_{,i}(\textbf{x}) = \frac{2i}{\beta} \mathcal{A}^1_{,i}(\textbf{x}) + \frac{ i}{\beta} \sum_{I=1}^{N_I} (\xi_I^0)_{i} \sum_{J = 1}^{N_{JI}} \Bigg\{ \sum_{m=1}^{\infty} \frac{(i \beta)^m}{m!} \left[ C^{m}_{2}(a_I, b_{JI}, l_{JI}^+) - C^{m}_{2}(a_I, b_{JI}, l_{JI}^-) \right] \\ & \quad \quad \qquad- a_I^2 \exp \left[ i \beta |a_I| \right] \left( \tan^{-1} \left[ \frac{l_{JI}^+}{b_{JI}} \right] - \tan^{-1} \left[ \frac{l_{JI}^-}{b_{JI}} \right] \right) \Bigg\}
\end{aligned}
    \label{eq:Helmholtz_stokes}
\end{equation} 
where
\begin{equation}
    C^{m}_{n}(a_I, b_{JI}, le) = \left(\sqrt{a_I^2 + b_{JI}^2}\right)^{m + n} \frac{le}{b_{JI}} F_1 \left[ \frac{1}{2}, \frac{-(m + n)}{2}, 1, \frac{3}{2}; \frac{-le^2}{a_I^2 + b_{JI}^2}, \frac{-le^2}{b_{JI}^2} \right]
\end{equation}
in which $F_1$ is defined in Eq. (\ref{eq:Appell}).

\section{Handling the mathematical singularity with time integral} \label{appendix:sing}
As Eq. (\ref{eq:3D_time}) indicates, when $n < m$ and $t = t_f$, there exist singularity issues. Following the established method in boundary element method \cite{Gupta1995}, let us interchange the integral sequence, and evaluate the time integral of the transient fundamental solution first: (multiplying by $(2 \alpha (t - t'))^n$): (n = 0, 1, 2 in this paper)

\begin{equation}
    \int_{t_{f-1}}^{t_f} (2 \alpha (t - t'))^{n} \thinspace G(\textbf{x}, \textbf{x}', t, t') \thinspace d t' = \frac{|\textbf{x} - \textbf{x}'|^{2 n - 1}}{2^{2 + n} \pi^{\frac{3}{2}} \alpha} \Big( \Gamma \left[ \frac{1-2n}{2}, \frac{|\textbf{x} - \textbf{x}'|^2}{4 \alpha (t - t_{f-1})} \right] - \Gamma \left[ \frac{1-2n}{2}, \frac{|\textbf{x} - \textbf{x}'|^2}{4 \alpha (t - t_f)} \right] \Big)
    \label{eq:int_seq_complete}
\end{equation}
where $\Gamma[a, b]$ is the upper incomplete Gamma function. When $(t - t_f) \rightarrow 0$ and $\frac{|\textbf{x} - \textbf{x}'|^2}{4 \alpha (t - t_{f-1})} \rightarrow \infty$, the incomplete Gamma function become zero: 

\begin{equation}
\begin{split}
    \lim_{t \rightarrow t_f} \int_{t_{f-1}}^{t_f} (2\alpha (t - t'))^n G(\textbf{x}, \textbf{x}', t, t') \thinspace d t' &= \frac{|\textbf{x} - \textbf{x}'|^{2n-1}}{2^{2 + n} \pi^{\frac{3}{2}} \alpha} \Big( \Gamma \left[ \frac{1}{2} - n, \frac{|\textbf{x} - \textbf{x}'|^2}{4 \alpha (t - t_{f-1})} \right] \Big) \\ & = \frac{|\textbf{x} - \textbf{x}'|^{2n-1}}{2^{2 + n} \pi^\frac{3}{2} \alpha} \left( \Gamma \left[ \frac{1}{2} -n \right] - \gamma \left[ \frac{1}{2} - n, \frac{|\textbf{x} - \textbf{x}'|^2}{4 \alpha (t - t_{f-1})} \right] \right)
\end{split}
    \label{eq:int_n_1}
\end{equation}
where $\gamma[a,b]$ is the lower incomplete Gamma function. Using the Taylor series expansion, the above equation can be expressed: 

\begin{equation}
    \lim_{t \rightarrow t_f} \int_{t_{f-1}}^{t_f} G(\textbf{x}, \textbf{x}', t, t') \thinspace d t' = \frac{|\textbf{x} - \textbf{x}'|^n}{2^{2+n} \pi^{\frac{3}{2}}\alpha} \Gamma\left[ \frac{1}{2} - n \right] - \frac{1}{ \pi^{\frac{3}{2}} \alpha} \sum_{m = 0}^{\infty} \frac{(-1)^m |\textbf{x} - \textbf{x}'|^{2 m}}{m! (2m + 1 - 2n) (4 \alpha (t_f - t_{f-1}))^{\frac{2m + 1 - 2n}{2}} } 
\end{equation}
where $\Gamma[\frac{1}{2}] = \sqrt{\pi}, \Gamma[-\frac{1}{2}] = -2 \sqrt{\pi}, \Gamma[-\frac{3}{2}] = \frac{4}{3} \sqrt{\pi}$. Subsequently, conducting the domain integral on $|\textbf{x} - \textbf{x}'|$ only, domain integrals yields as, 

\begin{equation}
    \begin{split}
    & \mathcal{C}^{n, f}(\textbf{x}, t) = \frac{(-2)^n}{\pi K (2n-1)!!} \int_{\Omega} |\textbf{x} - \textbf{x}'|^{2n-1} \thinspace d\textbf{x}' \sum_{m = 0}^{\infty} \frac{(2 \alpha)^n (-1)^m}{2 K (2 m + 1 - 2 n) (4 \alpha)^{m + \frac{1}{2}} \pi^{\frac{3}{2}} m!} (t - t_{f-1})^{\frac{1}{2} + n -m} \\ %
    & \quad \quad \quad \quad \quad \times  \sum_{I=1}^{N_I} \sum_{J = 1}^{N_{JI}} \frac{a_I b_{JI}}{2 (m + 1) (2 m + 3)} 
    \Big( A^m(a_I, b_{JI}, l_{JI}^+) - A^m(a_I, b_{JI}, l_{JI}^-) \Big)
    \end{split}
    \label{eq:3D_sing}
\end{equation}
where $!!$ refers to the double factorial. Note that when $t \rightarrow t_f$, it does not alter the second term in Eq. (\ref{eq:3D_time}), but the first term is time-independent. Specifically, when n = 0, the first term recovers the steady-state fundamental solution, $|\textbf{x} - \textbf{x}'|^{-1}$, namely the Newtonian potential; when n = 1, the first term yields $|\textbf{x} - \textbf{x}'|$, namely the biharmonic potential. These two domain integrals have been reported in our recent works \cite{Wu2021_polyhedral}. However, when n = 3, the first term becomes $|\textbf{x} - \textbf{x}'|^3$ can be evaluated as: 

\begin{equation}
\begin{split}
    \int_{\Omega} |\textbf{x} - \textbf{x}'|^3 \thinspace d\textbf{x}' & = \sum_{I=1}^{N_I} \sum_{J=1}^{N_{JI}} \left( \mathcal{P}(a_I, b_{JI}, l_{JI}^+) - \mathcal{P}(a_I, b_{JI}, l_{JI}^-) \right)
\end{split}
    \label{eq:psi_3}
\end{equation}
where 
\begin{equation}
\begin{split}
    \mathcal{P}(a_I, b_{JI}, le) & =  \frac{1}{240 a_I} \left[ 
    b_{JI} \sqrt{a_I^2 + b_{JI}^2 + \ell^2} \left(9a_I^2 + 5b_{JI}^2 + le^2\right) 
    \right. \\ & \left.+ 8a^4 \left(a_I - \sqrt{a_I^2}\right) \tan^{-1} \left(\frac{a_I + le - \sqrt{a_I^2 + b_{JI}^2 + le^2}}{b_{JI}}\right) 
    \right. \\ & \left.+ 8a^4 \left(a_I + \sqrt{a_I^2}\right) \tan^{-1} \left(\frac{a_I - le + \sqrt{a_I^2 + b_{JI}^2 + le^2}}{b_{JI}}\right) 
    \right. \\ & \left.- b_{JI} \left(15 a_I^4 + 10a^2b_{JI}^2 + 3b_{JI}^4\right) \log\left(-le + \sqrt{a_I^2 + b_{JI}^2 + le^2}\right)
\right]
\end{split}
\end{equation}

Alternatively, its first order partial derivative can be obtained using Stokes' theorem, 
\begin{equation}
\begin{split}
    \int_{\Omega} (|\textbf{x} - \textbf{x}'|^3)_{,i} \thinspace d\textbf{x}' & = \sum_{I=1}^{N_I} -(\xi_I^0)_{i} \sum_{J=1}^{N_{JI}} \left( \mathcal{Q}(a_I, b_{JI}, l_{JI}^+) - \mathcal{Q}(a_I, b_{JI}, l_{JI}^-) \right)
\end{split}
    \label{eq:psi_3_stokes}
\end{equation}
and 
\begin{equation}
    \begin{split}
    \mathcal{Q}(a_I, b_{JI}, le) & = \frac{1}{40} \left( b_{JI} le \sqrt{a_I^2 + b_{JI}^2 + le^2} (9 a_I^2 + 5 b_{JI}^2 + 2 le^2) 
    \right. \\ & \left. + 8 a_I^4 (a_I - \sqrt{a_I^2}) \tan^{-1} \left(\frac{a_I + le - \sqrt{a_I^2 + b_{JI}^2 + le^2}}{b_{JI}}\right) 
   \right. \\ & \left. + 8 a_I^4 (a_I + \sqrt{a_I^2}) \tan^{-1} \left(\frac{a_I - le + \sqrt{a_I^2 + b_{JI}^2 + le^2}}{b_{JI}}\right) 
    \right. \\ & \left.- b_{JI} (15 a_I^4 + 10 a_I^2 b_{JI}^2 + 3 b_{JI}^4) \log \left[-le + \sqrt{a_I^2 + b_{JI}^2 + le^2}\right] \right)
    \end{split}
\end{equation}

\section{Convergence on series expression} \label{appendix:con}
\begin{figure}
    \centering
    \includegraphics[width = 1.0 \linewidth, keepaspectratio]{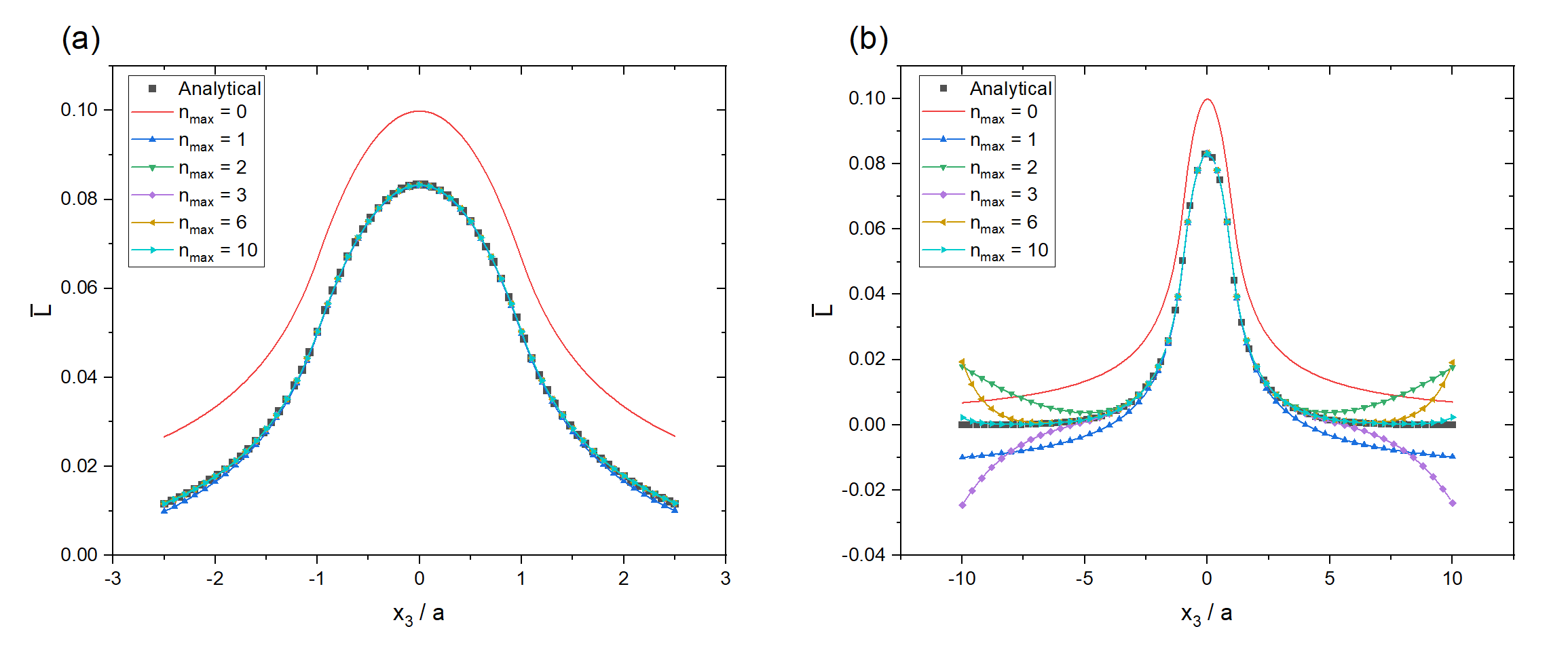}
    \caption{Comparison of number of series $n_{max}$ (0, 1, 2, 3, 6, 10) of time and domain integrals of the transient fundamental solution $\overline{L}$ over a spherical inclusion $a = 0.1$ m along the vertical centerline (a) $x_3 \in [-2.5, 2.5] a$ and (b) $x_3 \in [-10, 10] a$, when the sphere is approximated by $N_I = 3, 664$-surface polyhedron.}
    \label{fig:convergence}
\end{figure}

This appendix illustrates the convergence of series expansion on domain integrals of the transient fundamental solution. Note that thermal properties and inclusion radius are retained from Section 4. (a). To validate its convergence, this subsection follows the example of Section 5.1, where $\alpha = 0.05$ and $t_i = 0, t_f = t = 2 $s. Since $t_f = t$, the revised formulae as Eq. (\ref{eq:3D_sing}) is applied. Following the numerical example, the sphere is approximated by $3, 664$ quadrilateral surface elements and the number of series (implemented) $m = 0, 1, 2, 3, 6$ and $10$. Shown in Fig. \ref{fig:convergence} (a), when the field points are within a shorter range, i.e., $|x_3| \leq 2.5 a$, each case can provide good results except $n_{max} = 0$. As $x_3$ approaches $2.5 a$, the case $n_{max} = 1$ starts to exhibit greater numerical errors, which implies that when the distance becomes larger, the approximation requires the involvement of more series terms. To verify this, Fig. \ref{fig:convergence} (b) plots the variation of approximated tensor $\overline{L}$ when field points are further away. It can be observed that when $x_3 \geq 3.5 a, 5 a, 7.5 a$, the curve $n_{max} = 2$, $n_{max} = 3$ and $n_{max} = 6$ starts to exhibit obvious approximation errors, respectively. Although the approximation accuracy decreases with increasing distance, $n_{max} = 10$ can provide results accurate enough for engineering applications within ten times the radius.

\section{Verification of linear, quadratic time Eshelby's tensors}
\label{appendix:eshel}
This appendix section completes the verification for linear and quadratic time Eshelby's tensors. Since there only exists closed-form solutions for spherical inclusions, this section will compare numerical results with transition functions Eq. (\ref{eq:sphere_time_lin}) and Eq. (\ref{eq:sphere_time_qua}), and then verify Eq. (\ref{eq:stokes_time}) (partial derivatives) to generate time Eshelby's tensors. Based on the comparison in Section 5.1.2, the approximation by $3,664$ polyhedrons is applied, while other settings (properties, time) remain the same. 

% verification of e1 and e2
\begin{figure}
\centering
\includegraphics[width = \textwidth, keepaspectratio]{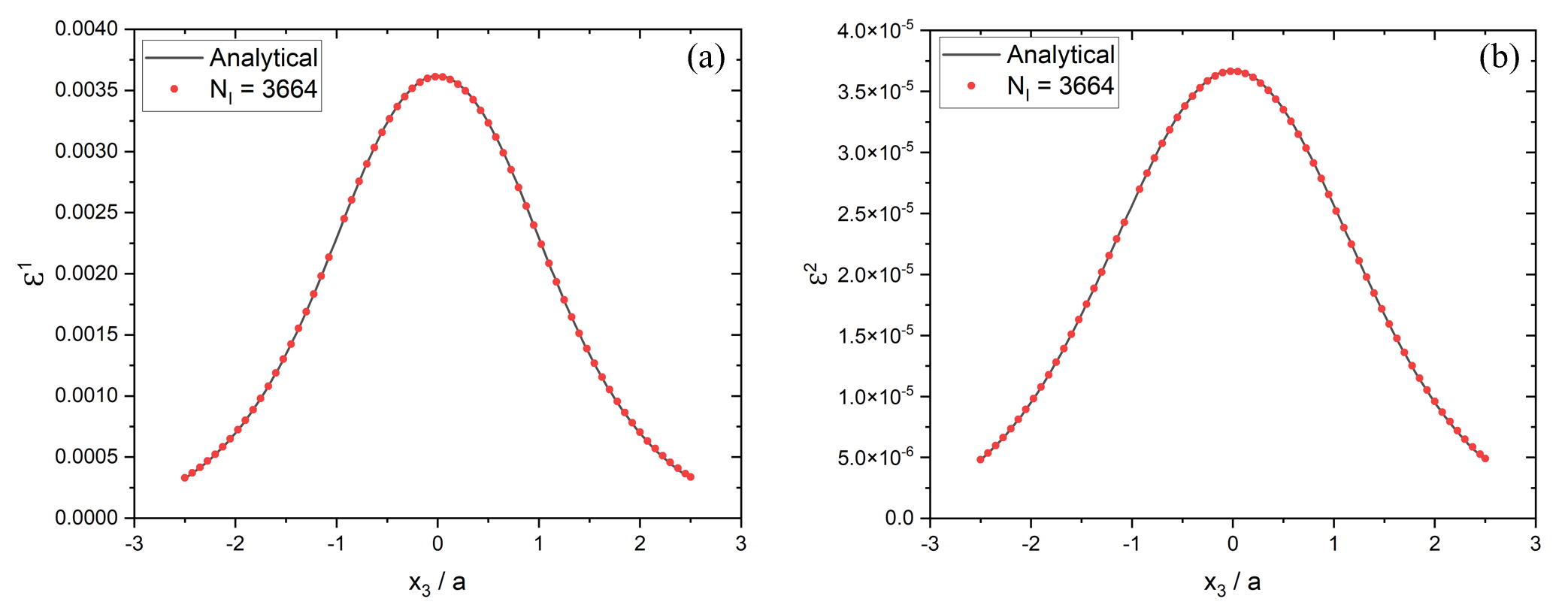}
\caption{Reproduction and verification of the analytical spherical integral with radius $a = 0.1$ m containing a uniform heat source (existing within [0, 2] s), along the vertical center line $x_3 \in [-2.5, 2.5] a$ on (a) analytical expression $\mathcal{E}^1$ in Eq. (\ref{eq:sphere_time_lin}); and (b) analytical expression $\mathcal{E}^2$ in Eq. (\ref{eq:sphere_time_qua}), when $3,664$-surface polyhedron is applied.}
\label{fig:e1_e2}
\end{figure}

\begin{figure}
\centering
\includegraphics[width = \textwidth, keepaspectratio]{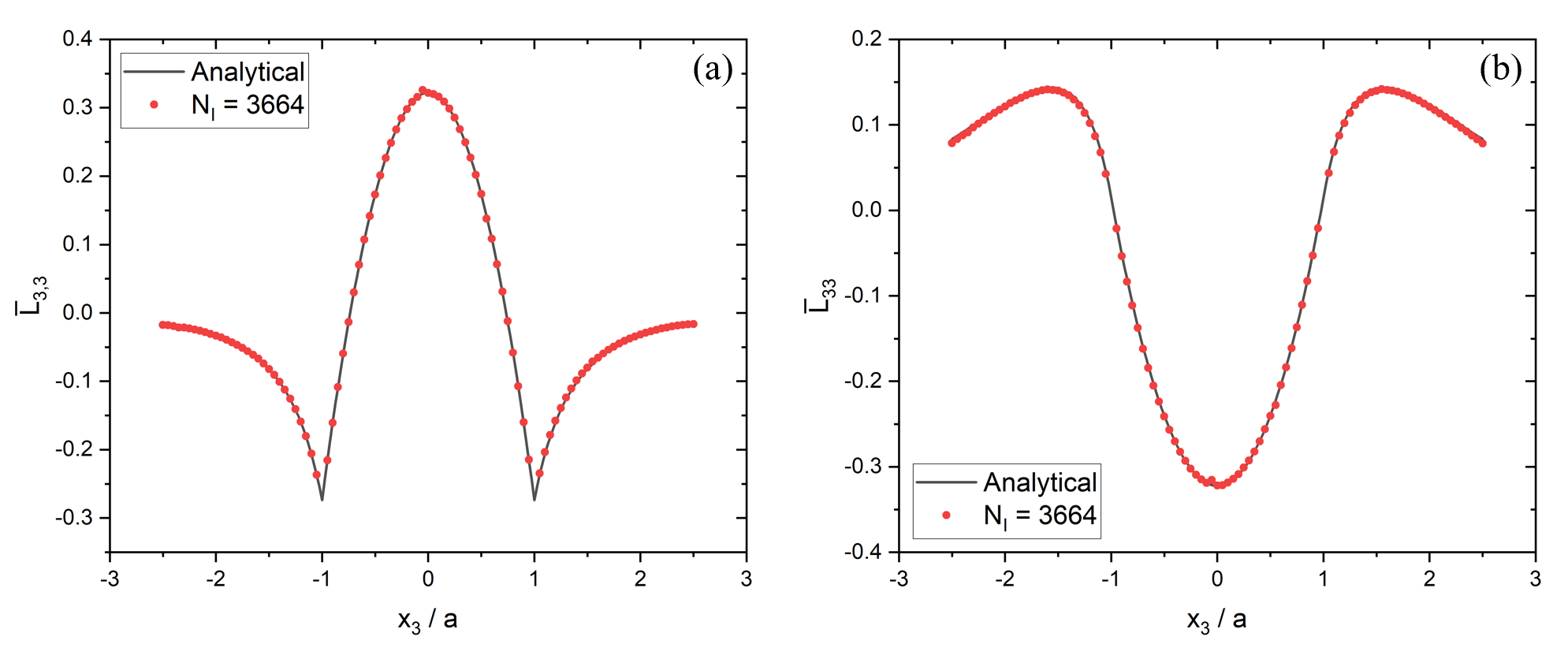}
\caption{Reproduction and verification of the analytical spherical integral with radius $a = 0.1$ m containing a uniform heat source (existing within [0, 2] s), along the vertical center line $x_3 \in [-2.5, 2.5] a$ on (a) linear Eshelby's tensor $\overline{L}_{3,3}$; and (b) quadratic Eshelby's tensor $\overline{L}_{33}$, when $3,664$-surface polyhedron is applied.}
\label{fig:lin_qua_tensors}
\end{figure}

As Figs. \ref{fig:e1_e2} (a-b) indicate, the numerical approximation by Eq. (\ref{eq:3D_time}) agree well with analytical expressions in Eq. (\ref{eq:sphere_time_lin}) and Eq. (\ref{eq:sphere_time_qua}). In the following, Figs. \ref{fig:lin_qua_tensors} (a-b) show variation of linear Eshelby's tensor $\overline{L}_{3,3}$ and quadratic Eshelby's tensor $\overline{L}_{33}$, respectively. For thorough verification, $\overline{L}_{3,3}$ was chosen over $\overline{L}_{3}$ because it involves the second-order partial derivative of the elementary function $\mathcal{E}^{1}$, therefore validating the partial differentiation chain rule. Simultaneously, $\overline{L}_{33}$ inherently includes the second-order partial derivatives of $\mathcal{E}^{2}$, which further confirms the consistency of the formulation for Eshelby's tensor with polynomial-form eigen-fields. 

% verification of e1 and e2

\FloatBarrier

\bibliographystyle{unsrt}

\end{document}